\documentclass[trackchanges,twocolumn]{aastex701}

\graphicspath{{../figures/}{../figures/}}

\begin{document}

\title{A Modified Multi-Level Tracking Scheme for the Detection of Sunspot Umbral Dots}%

\author[orcid=0000-0001-5963-8293,sname='Louis']{Rohan Eugene Louis}
\affiliation{Udaipur Solar Observatory, Physical Research Laboratory, Dewali Badi Road, Udaipur --313001, Rajasthan, India}
\email[show]{rlouis@prl.res.in}  
\correspondingauthor{Rohan Eugene Louis}

\author[orcid=0009-0003-6122-8623,sname='Chaturvedi']{Amit Chaturvedi} 
\affiliation{Udaipur Solar Observatory, Physical Research Laboratory, Dewali Badi Road, Udaipur --313001, Rajasthan, India}
\email{amitchaturvedi@prl.res.in}


\begin{abstract}
Umbral dots (UDs) are small-scale convective intrusions in the umbral core of sunspots and pores. Different methods have been used in the past to determine 
the physical properties of UDs. One of the methods typically used is multi-level tracking (MLT), which tags spatial structures at equi-spaced intensity levels from the  
highest level while progressing downward. A modified approach to the regular MLT is explored in this article that also uses the local 
intensity maxima with a change in the threshold condition to enclose a UD, such that diffuse UDs do not appear extended than they visually appear. 
The physical properties of UDs from these two MLT approaches are compared. The methods are implemented on high-resolution blue continuum images of four sunspots 
from the 50-cm Solar Optical Telescope on board \textit{Hinode}. In addition, we introduce a density-based, spatial clustering routine for the first 
time to ascertain the differences resulting from the two tracking methods. 
The modified MLT approach yields an effective diameter with median values ranging from 250--310\,km which is on average 70--90\,km smaller than the regular
MLT approach. The lower effective diameter in the modified method is associated with a reduced UD fill fraction of 12\%--13\% while the regular method yields 17--19\%.
However, these differences are still within the range of values cited by earlier works. On the other hand, the histogram of 
the mean intensity of UDs from both methods is nearly identical. The spatial clustering of UDs from both methods also shows very similar results. 
There is, however, a preferential spatial concentration of UDs, particularly at locations where the umbral core is highly irregular and in the vicinity of 
faint light bridges. The dependency of the localized clustering of UDs on the evolutionary phase of the sunspot and its magnetic complexity
needs to be further explored.
\end{abstract}

\keywords{\uat{Sunspots}{1653} --- \uat{Solar Magnetic Fields}{1503} --- \uat{Solar Photosphere}{1518}}



\section{Introduction}
\label{intro}
Sunspots are locations of intense magnetic fields where the photospheric magnetic field strength is in excess of 2.5\,kG. The presence of
strong magnetic fields suppresses overturning convective motions within sunspots \citep{1941VAG....76..194B,1963MNRAS.126..431C}, rendering their 
appearance to be darker than the surrounding quiet-Sun (QS). While the umbra is the darkest region of a sunspot, the emitted 
energy is still 10--20\% of that in the QS, which cannot be carried solely by radiation. This surplus brightness is the result 
of convective motions that are greatly reduced in the presence of strong magnetic fields \citep{1965ApJ...141..548D}, and is 
associated with umbral dots (UDs) -- sub-arcsecond, bright features that populate the dark umbral background 
\citep{1964ApJ...139...45D,1968SoPh....4..303B,1979A&A....79..128L}. 

UDs can be considered as either intrusions of hot, non-magnetized plasma within the ``gappy'' umbral magnetic field 
\citep{1979ApJ...234..333P,1986ApJ...302..809C}, or oscillatory convection in thin columns of a monolithic sunspot \citep{1990MNRAS.245..434W}.
Three-dimensional radiative magneto-hydrodynamic (MHD) simulations of magneto-convection by \cite{2006ApJ...641L..73S} depict UDs as narrow, 
convective plumes, which radiatively cool near the continuum-forming height. The UDs appear as elongated structures, with a central dark lane 
that coincides with strong upflows that are flanked by relatively weaker downflows. The fine structure in UDs, as predicted by numerical 
simulations, has been observationally verified and described in \cite{2008ApJ...672..684R,2009A&A...504..575S,2010ApJ...713.1282O}.

UDs are generally categorized as ``central'' and ``peripheral'' \citep{1986A&A...156..347G}
structures based on their relative location in the umbra and have intensities of about 
0.34\,$I_{\textrm{\tiny{QS}}}$ and 0.48\,$I_{\textrm{\tiny{QS}}}$, respectively \citep{1997A&A...328..689S,2002A&A...388.1048T}, while 
\cite{2008A&A...492..233R} reported values of 0.53\,$I_{\textrm{\tiny{QS}}}$ and 0.65\,$I_{\textrm{\tiny{QS}}}$, respectively.
On the other hand, the ratio of the peak-to-background intensity is about $1.15$ for central UDs and and about $1.23$ for peripheral
ones \citep{2008A&A...492..233R}, while \cite{2012ApJ...757...49W} reported values of about $1.46$ and $1.5$, respectively. In general,
larger and long-lived UDs are seen in areas of enhanced background intensity \citep{1997A&A...328..682S}. The sizes of UDs can range from 
125\,km to 400\,km, as reported by various authors using different identification routines 
\citep{1997A&A...328..682S,2002A&A...388.1048T,2009A&A...504..575S,2009ApJ...694.1080S,2009ApJ...702.1048W,2012ApJ...752..109L,2012ApJ...757...49W}.
UDs typically fill around 6--18\% of the umbral area \citep{1997A&A...328..682S,2002A&A...388.1048T,2018ApJ...855....8Y}.

The differences in values of the physical properties can be attributed to the various techniques utilized in the 
identification of UDs as well as the type of sunspot analyzed. These detection methods include, edge enhancement with 
thresholding \citep{1997A&A...328..682S}, low-noise curvature detection \citep{2008SoPh..250...17H,2009A&A...504..575S}, 
intensity thresholding \citep{2009ApJ...702.1048W}, and multi-level tracking 
\citep[MLT;][]{2008A&A...492..233R,2010A&A...510A..12B,2012ApJ...752..109L,2018ApJ...855....8Y}. In this article, we 
implement MLT and a modified version of it, which uses the local intensity maxima and an altered threshold, 
on a set of four sunspots to determine the differences in the size, peak-to-background ratio, and fill fraction. We also introduce a 
density-based clustering routine for the first time to sort UDs on the basis of their spatial positions and ascertain the differences 
arising from the two tracking schemes.   
The rest of the article is organized as follows: 
Sect.~\ref{obs} summarizes the observations,  while Sect.~\ref{analyse} describes the detection and clustering techniques. The results are presented 
in Sect.~\ref{results}. The discussion and conclusions are presented in Sects.~\ref{discuss} and \ref{conclu}, respectively.

\section{Observations}
\label{obs}

\begin{figure*}[!ht] 
\centerline{
\hspace{45pt}
	\begin{minipage}{0.8\textwidth}
	\centering
	\includegraphics[angle = 90,width=\textwidth]{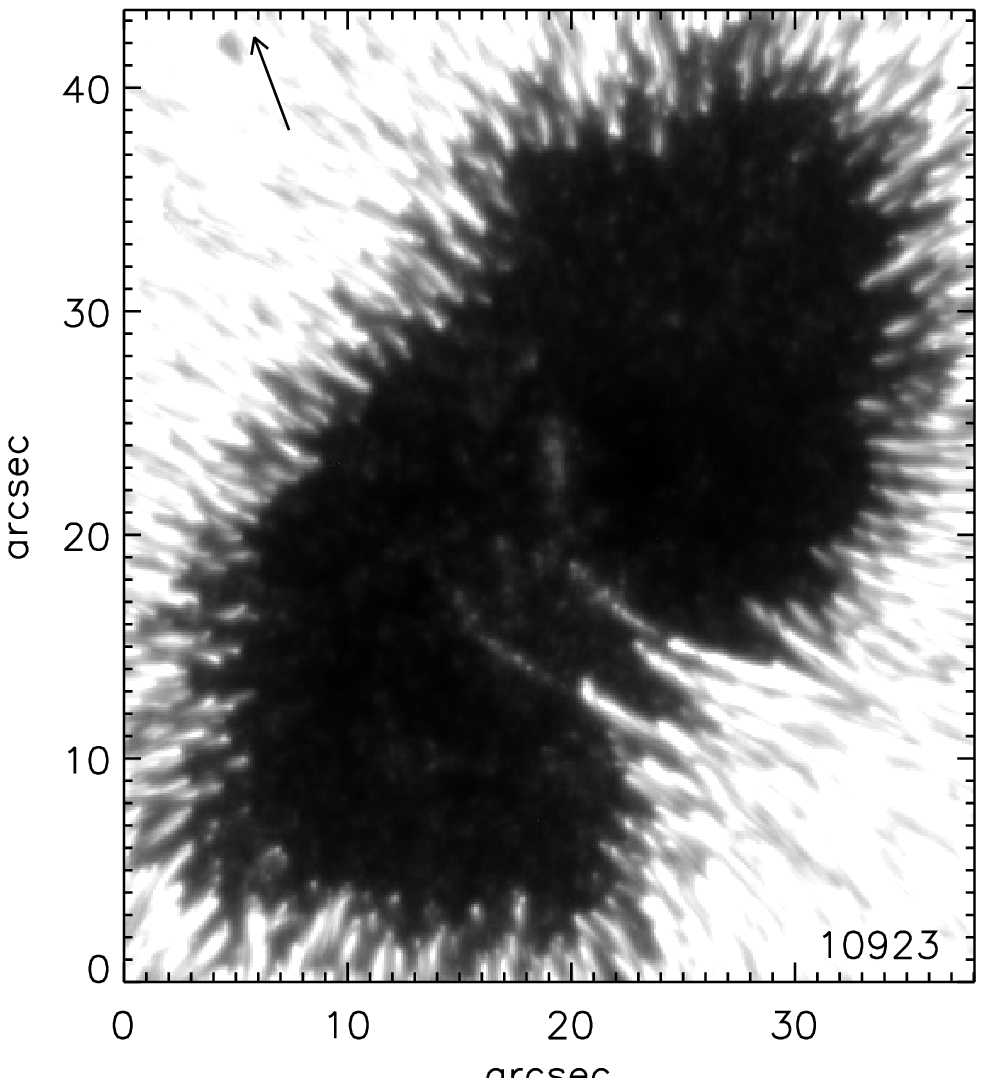}
	\end{minipage}
	\hspace{-160pt}
	\begin{minipage}{0.55\textwidth}
	\centering
	\includegraphics[angle = 90,width=\textwidth]{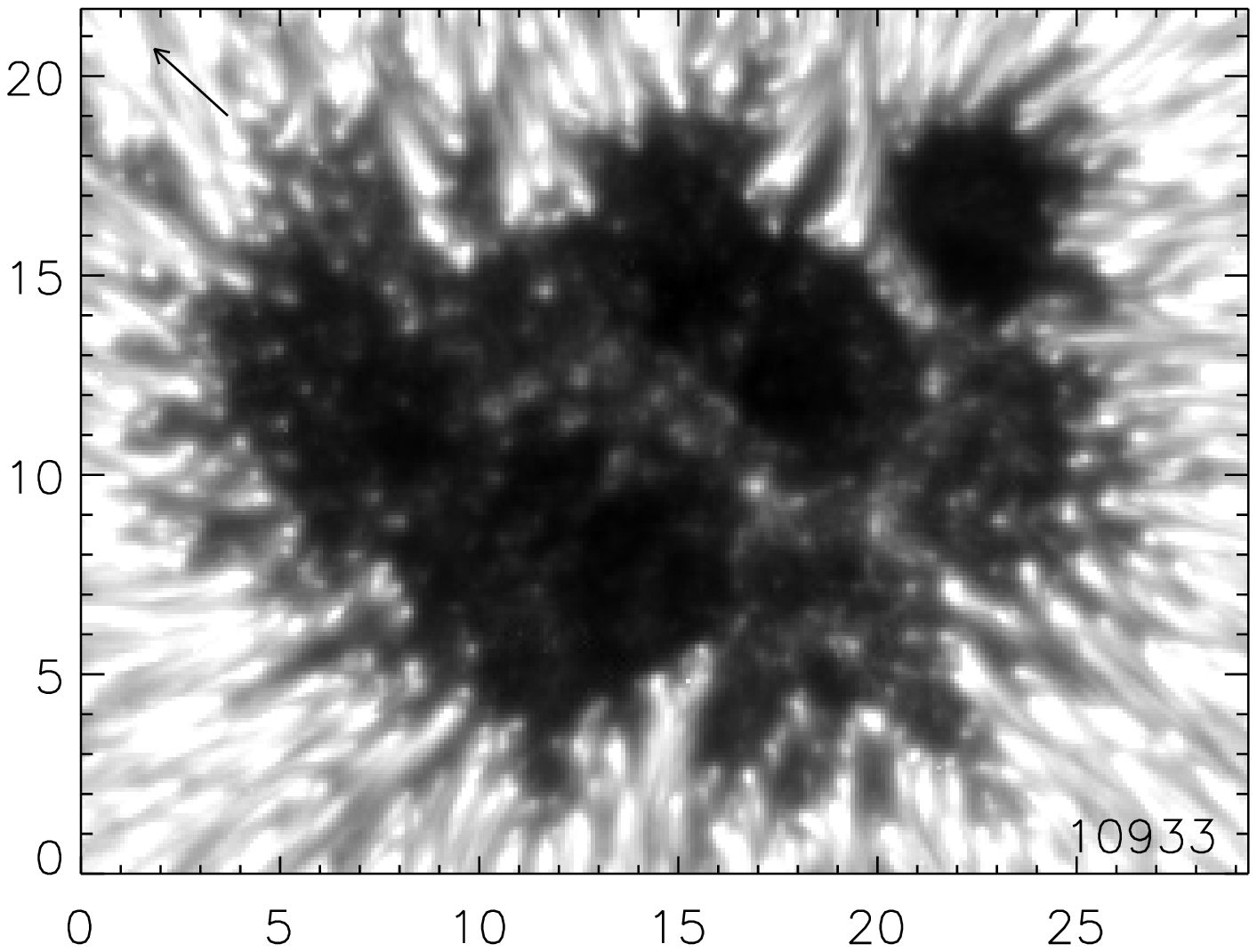}
	\end{minipage}
}
\vspace{-20pt}
\hspace{5pt}
\centerline{
	\begin{minipage}{0.78\textwidth}
	\centering
	\includegraphics[angle = 90,width=\textwidth]{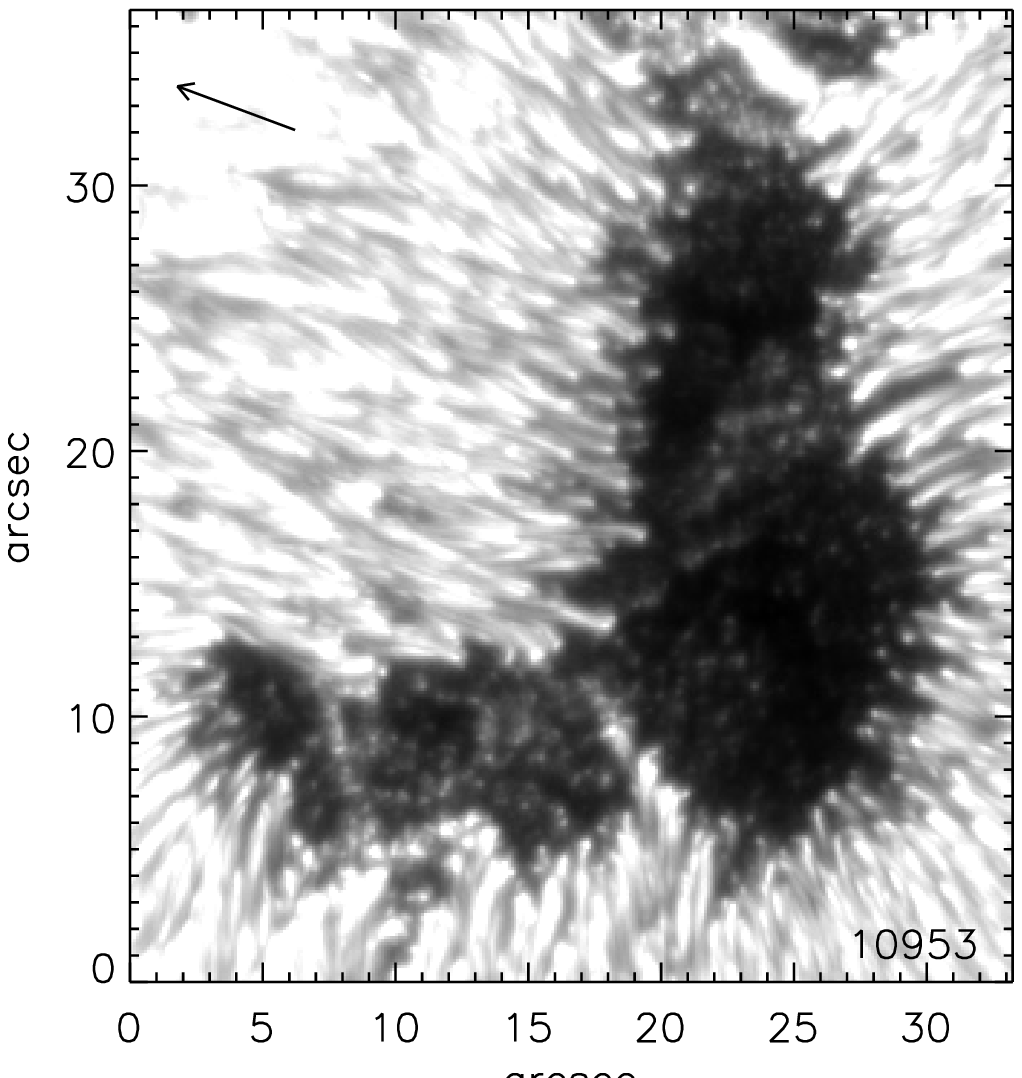}
	\end{minipage}
	\hspace{-140pt}
	\begin{minipage}{0.48\textwidth}
	\centering
	\includegraphics[angle = 90,width=\textwidth]{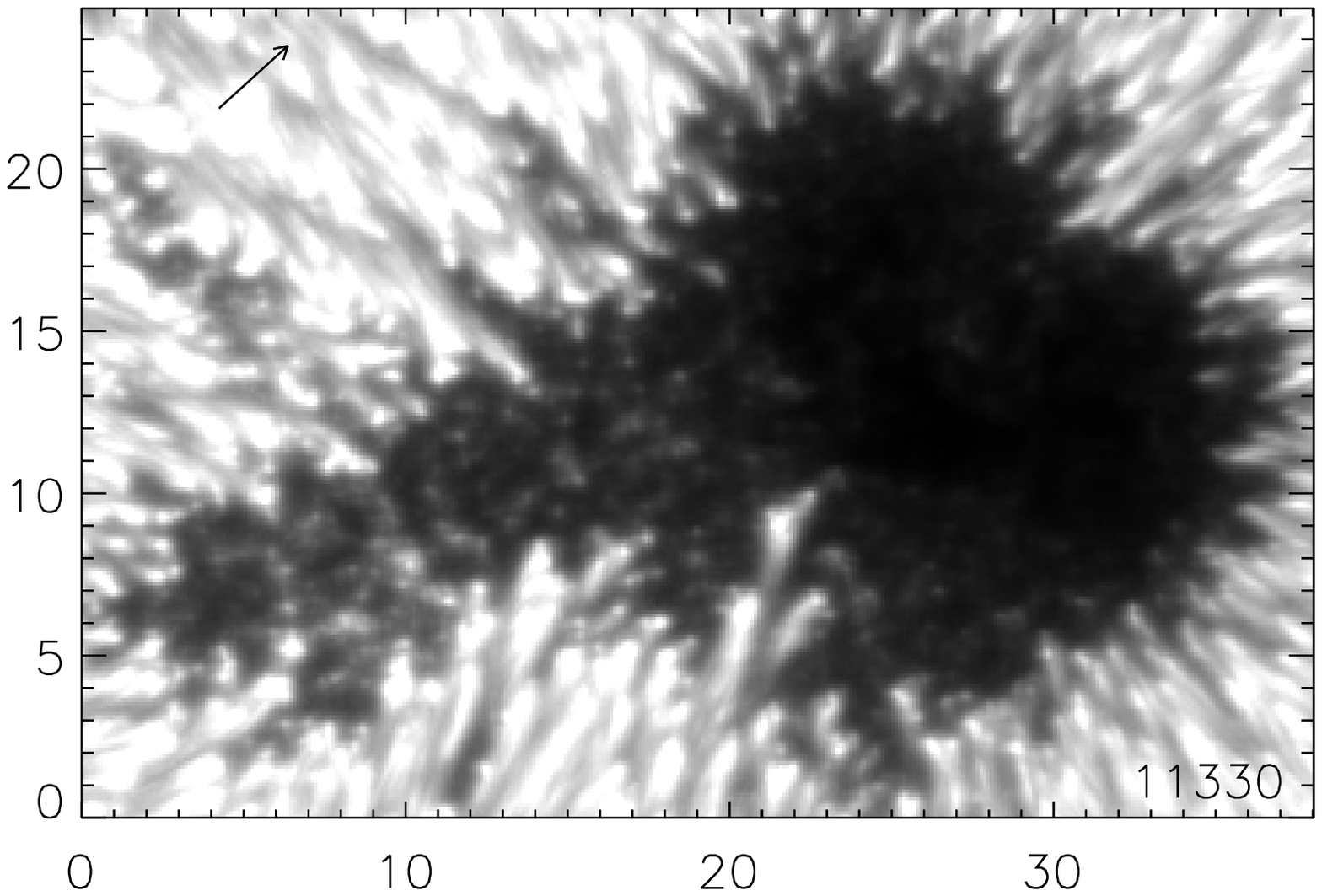}
	\end{minipage}
}
\vspace{0pt}
\caption{Sunspot umbrae analyzed in this work. The NOAA AR is inscribed in the bottom-right corner, while
the arrow in the top-left corner points to disk center. The images have been clipped in intensity to 
show the detail within the umbra.}
\label{fig01}
\end{figure*}

We utilize high-resolution blue continuum filtergrams from the 50-cm Solar Optical Telescope \citep{2008SoPh..249..167T} on board the Japanese
satellite \textit{Hinode} \citep{2007SoPh..243....3K}. The filtergrams have a spatial sampling of 0\farcs109 and were taken with a broad-band 
filter with a width of 4\,\AA\, and centered at 450.45\,nm. The four sunspots analyzed in this study are NOAA AR 10923, 10933, 10953, and 11330, which 
were observed by \textit{Hinode} on 2006 November 14, 2007 January 05, 2007 May 02, and 2011 October 27, respectively. The four sunspots were located at a 
heliocentric angle of $8^{\circ}$, $2^{\circ}$, $16^{\circ}$, and $2^{\circ}$, respectively.
The Level-0 data were processed to Level-1 data using the \textit{``fg\_prep''} routine in SolarSoft, which carried out dark correction, flat-fielding, 
and removal of bad pixels. The average QS was then used to normalize the images. 
To extract the umbra-penumbra boundary, we used the Otsu's histogram shape-based, image thresholding technique \citep{otsu_imgseg,liao2001fast}
and smoothed the resulting contour by 11 pixels. 
Even though UDs are highly dynamic, we analyzed a single image for each sunspot as we were interested in the differences from the 
two detection schemes. The use of multiple sunspots ensured that the distinct morphology, evolutionary epoch, and area of the four active regions
rendered the analysis equivalent to a frame by frame comparison for a time series dataset. 
Figure~\ref{fig01} shows the sunspots analyzed in this article.

\section{Data Analysis}
\label{analyse}

\begin{figure*}[!ht] 
\centerline{
\includegraphics[angle = 90,width=\textwidth]{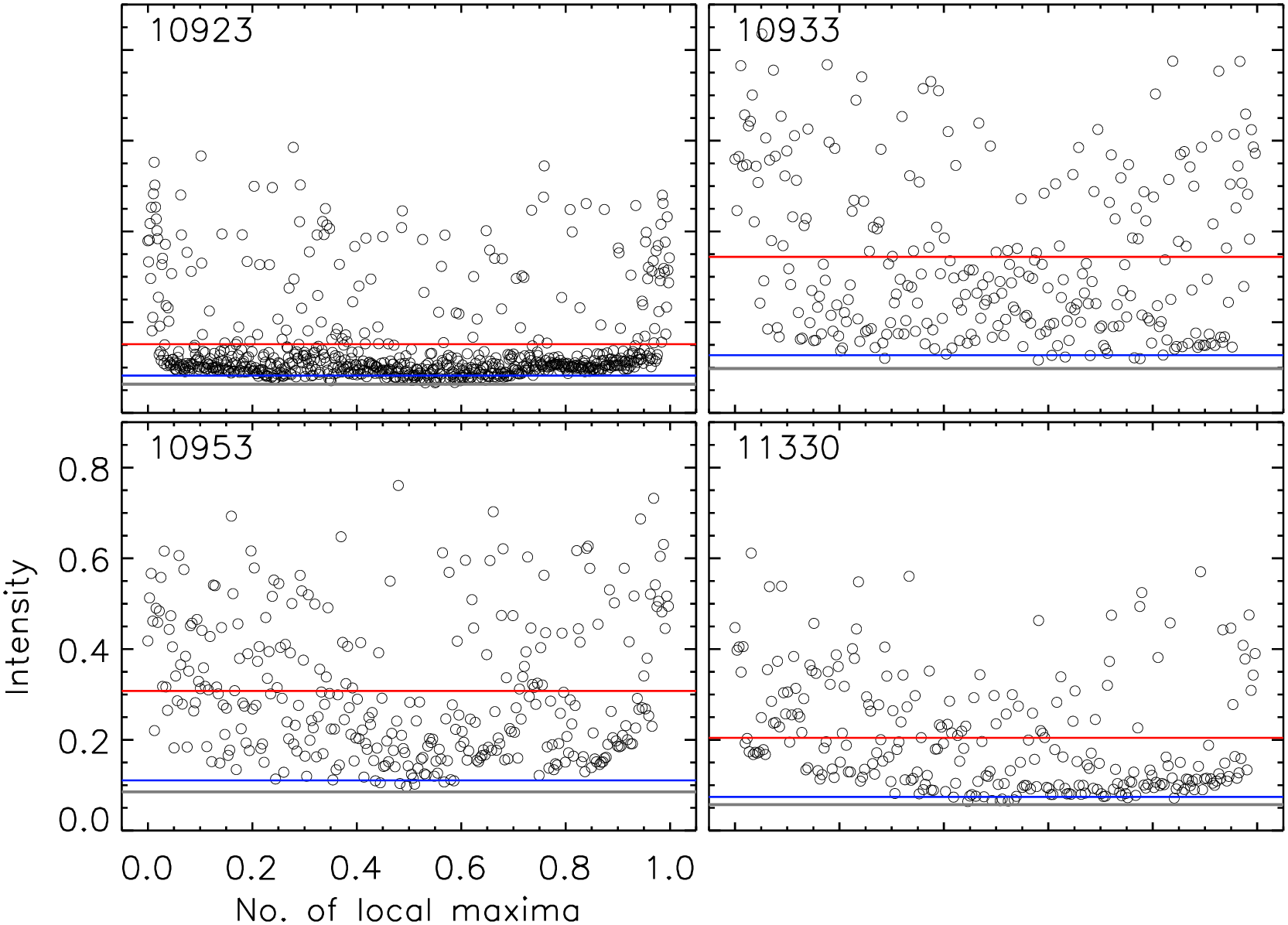}
}
\vspace{0pt}
\caption{The intensity at the locations of the local maximum within the umbra shown with open circles.
The gray line depicts the minimum intensity level ($I_{\textrm{\tiny{min}}}$) within the umbra, the blue line 
at $1.3I_{\textrm{\tiny{min}}}$, and the red line marks the mean value of intensity greater than $1.3I_{\textrm{\tiny{min}}}$.}
\label{fig02}
\end{figure*}

\subsection{Multi-Level Tracking (MLT)}
\label{mlt}
MLT \citep{2001SoPh..201...13B} is a method that involves splitting the intensity levels in an image into equi-spaced 
values and tracking structures from the top-most intensity level to the bottom-most level, tagging features as they appear at each level to either
previously detected structures or new features that appear for the first time at that level. The one-dimensional and two-dimensional illustrations of MLT
are described in \cite{2008A&A...492..233R} and \cite{2012ApJ...752..109L}, respectively. Prior to implementing MLT on each sunspot image, we determined
the local intensity maximum and minimum within a 5$\times$5\,pixel, neighbourhood, whose values are shown in Fig.~\ref{fig02}. The general 
distribution of intensity values shows that 60\% of the points (78\% in the case of AR 10923) are faint and diffuse. The minimum intensity 
within the umbra is indicated by the gray line, while the blue line is 30\% higher than the minimum value and is chosen as the new minimum intensity 
level in the umbra for MLT. The reason for choosing a 30\% threshold is to avoid spurious or noisy structures from
being identified by MLT, as lower intensities correspond to the darkest regions of the umbra that are generally devoid of any features that can be 
resolved. In addition, the local maxima detection routine picks up points that have the highest value in the neighborhood, irrespective of the intrinsic
intensity of that patch in the umbra. The mean value of all local intensity maxima, exceeding the minimum value, is indicated by
the red line. The range of the mean values lie between 0.15 in AR 10923 to 0.35 in AR 10933. We choose the pixels of local intensity maxima that have 
values between the red and the blue lines in the figure and save their spatial position.
Once MLT completes indexing all features in the umbra, the selection of the feature in each indexed patch is done using the mean of the maximum 
intensity and background intensity, i.e., $0.5*(I_{\textrm{\tiny{max}}} + I_{\textrm{\tiny{bck}}})$, 
which yields a contour around the feature. The background intensity is an image that is devoid of UDs and is determined using the procedure 
described in \cite{1993PatRe..26..375B}.
The area enclosed by the contour is calculated with the boundary points using the `shoelace method'
\citep{1769Meister}, which is based on the `triangle method', wherein $A=0.5|\sum_{i=1}^{n} (x_i y_{i+1} - x_{i+1} y_i)|$, with ($x_{n+1}, y_{n+1}) = (x_1, y_1)$.
The area within the contour should exceed four pixels, and the contour should have a two-pixel separation from the 
umbra-penumbra boundary. The corresponding contour outlining the feature is indexed, labeled ``A'', and the coordinates saved. The region enclosed
by the contour is set to one in a binary map, which is repeated for each indexed region detected by MLT. If the contour encloses points of local maxima, 
then the contour label is changed to ``B''. We utilize the ``ray-casting'' \footnote{https://www.youtube.com/watch?v=TA8XQgiao4M}
procedure to determine if our point of local maximum is within the contour or not.
A separate list is created for those local maxima points that do not have any contour enclosing them. 
The reason the residual local maxima points lack a contour around them is due to the threshold of the mean value between the maximum and background 
intensity used by MLT. In an additional step, hereafter referred to as stage-2, we center a circle on the point having the local maximum with a 3-pixel 
radius and determine which pixels are assigned a value of zero from the previously created binary image for the MLT contours. 
The contour is determined around the local maximum point using a level equal to the mean value between the maximum and minimum intensity and is saved if the 
area is greater than four pixels. The contours determined in stage-2 are assigned a label ``C''.

In the modified scheme, we implement MLT as before, but we check if the patch contains points of local maximum. If there are no points of local maximum, then the
contour is saved and labeled ``A'' as in the regular case, ensuring that the area exceeds 4 pixels. If the patch contains points of local maximum, then 
all contours within the patch, having a threshold exceeding 50\% of the sum of maximum intensity and mean intensity in the patch, 
i.e., $0.5*(I_{\textrm{\tiny{max}}} + I_{\textrm{\tiny{mean}}})$,
are determined and saved if
their area exceeds 4 pixels. This set of contours is labeled ``B''. The reason for 
using a different threshold in the modified approach is to prevent diffuse or faint structures to become more extended than they appear visually.
The above process is repeated until all the patch indices from MLT have been checked. 
The remaining points of local maximum that are not enclosed by a contour are subjected to stage-2 processing, as described earlier for the regular MLT scheme.
To summarize, the regular and modified schemes differ in the manner in which ``B''-labeled contours are determined, 
with distinct threshold values for both methods that produce a single contour in the former approach and multiple contours in the latter. 

\subsection{Clustering of UDs}
\label{dbscan} 
While several clustering routines are available for sorting multi-dimensional datasets, \textit{k-means} is the most widely used unsupervised scheme, particularly 
in solar physics for grouping spectral profiles \citep{2011A&A...530A..14V,2019ApJ...875L..18S,2024AdSpR..73.3256L}. Given the number of clusters, \textit{k-means} 
utilizes Euclidean distances between the data points and a set of initial random cluster centers to assign each point to a cluster based on the minimum distance. 
At the end of each iteration, the centroids of each cluster are updated, and the sorting continues until the cluster centroid and the variance of each cluster stabilize. 
The advantages of \textit{k-means} are its simplicity, speed, and effectiveness to handle large, high-dimensional datasets. However, given its centroid-based 
implementation, it fails in situations where there are outliers in the data and if the clusters have different sizes or densities. This is particularly seen in 
the sunspot umbral core, where either one or all the following conditions may exist, such as the absence of UDs in the dark nuclei of the umbra, clumps of UDs 
appearing close to the umbra-penumbra boundary, or large-scale clustering of UDs forming faint light bridges. In this article, we utilize a Density Based Spatial
Clustering of Applications with Noise (DBSCAN) for grouping UDs that were detected from the two MLT schemes using the centroid positions derived from the various 
contour labels. DBSCAN can deal with outliers, datasets with varying densities, and does not require a number of clusters to sort the data into, like its counterpart
\textit{k-means}. However, it requires two parameters, namely, the radius or eps circle ($\epsilon$) and the min points ($M$). The clustering works by segregating 
points into core, border, and noise points in the following manner. A core point is one that contains points greater than or equal to $M$ within  
a circle of radius $\epsilon$ centered on it. 
All other points that do not satisfy the above condition are labeled noise points initially. In order to assign a cluster number to all the 
core points and separate the border points from noise points, the first core point is assigned an index one, and all points within its $\epsilon$ circle are also 
assigned the same index. We will refer to this as Step-1. 
A separate list is created with those points within this $\epsilon$ circle that are assigned an index one and checked one-by-one if they are core points or border 
points. We will refer to this as Step-2. If a core point is encountered, i.e., there are points greater than or equal to $M$ in the $\epsilon$ circle, then these 
additional points are assigned an index one and added to the list 
described above. If there are less than $M$ points, the point is a border point. It is labeled as such, but no further points in its $\epsilon$ circle are added 
to the list. In either case, a flag is set to indicate that this point has been verified and the program will exclude it when accessing the list containing points with
an index one. Step-2 is repeated until all the points in the list with an index one, which are yet to be verified, have been exhausted. The index is then incremented 
to two, and Step-1 is repeated over all remaining core points that have not been assigned an index. In this manner, one obtains several clusters consisting of core 
and border points, as well as a number of noise points/outliers. This entire routine was written in the Interactive Data Language
\footnote{https://github.com/Rohan-Louis-81/IDL} and verified with the standard
``DBSCAN'' function in Python, which is part of the ``Scikit-learn'' library.

\begin{figure*}[!ht] 
\centerline{
\hspace{180pt}
\includegraphics[angle = 90,width=0.9\textwidth]{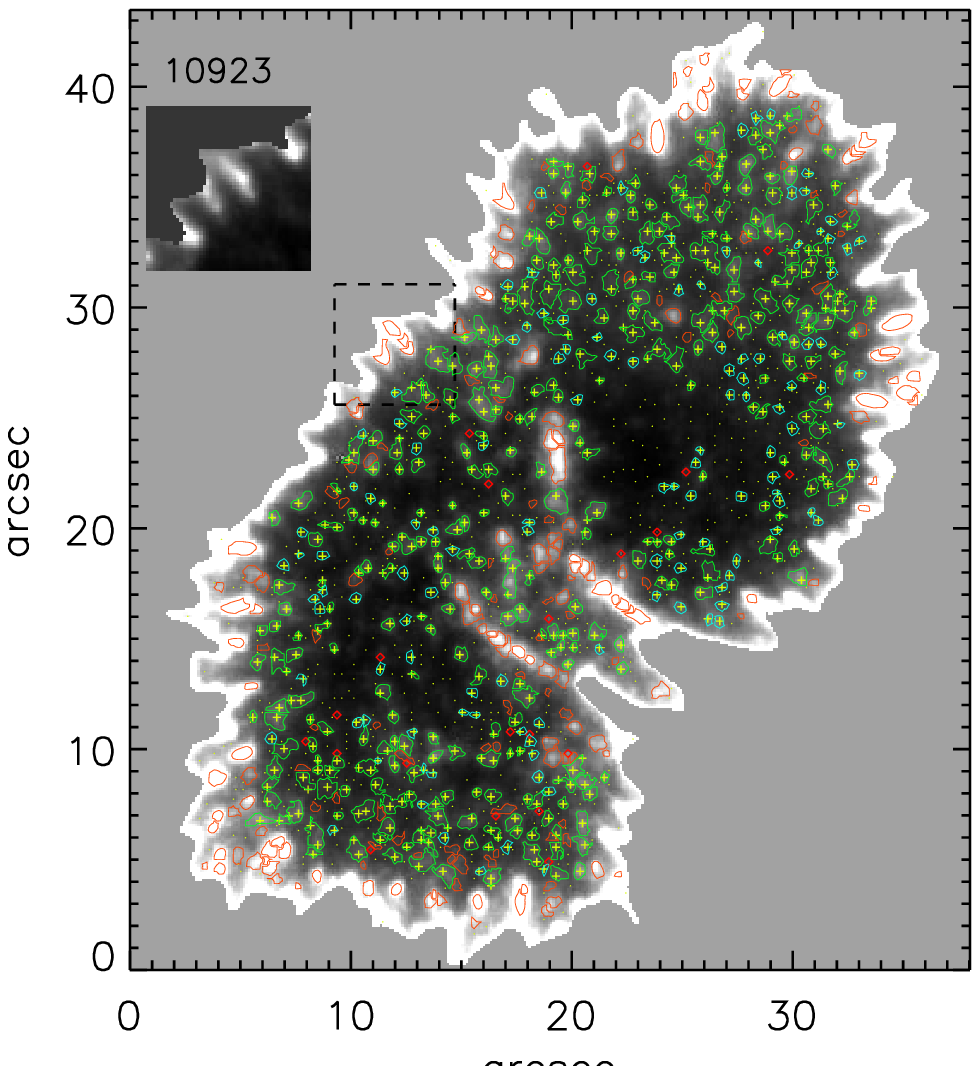}
\hspace{-220pt}
\includegraphics[angle = 90,width=0.9\textwidth]{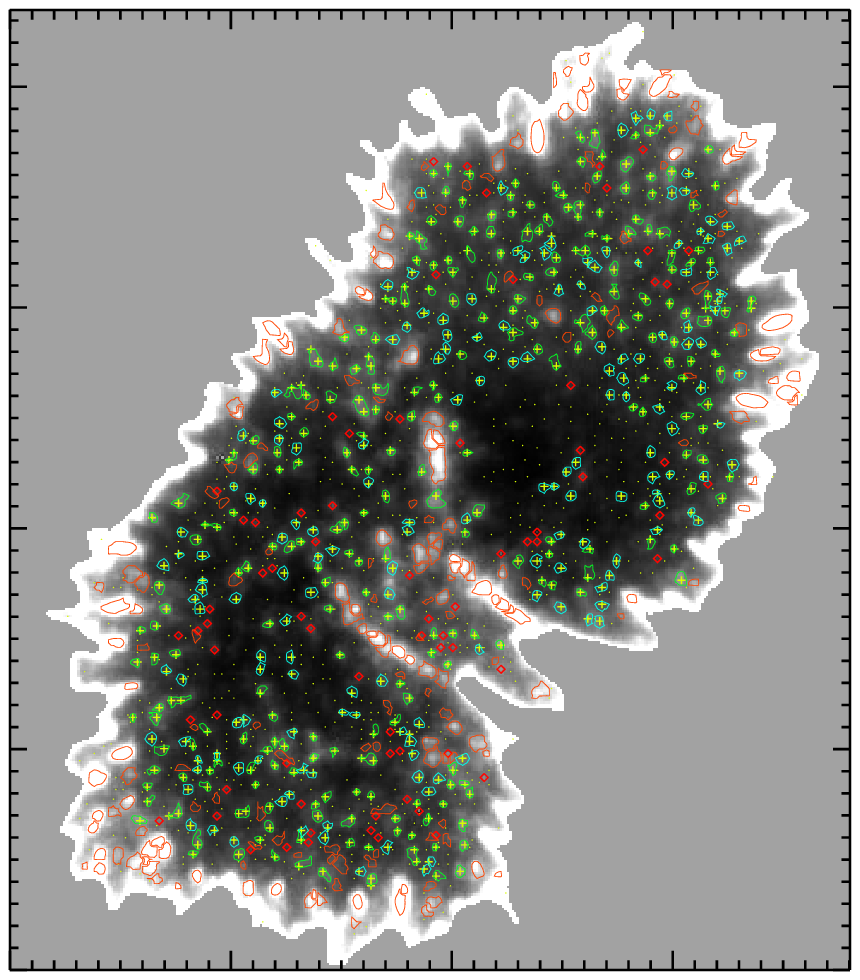}
}
\vspace{5pt}
\caption{Identification of UDs using the MLT scheme for AR 10923. The red contours outline UDs that do not contain a 
pixel with a local maxima. The green contours are those that contain a local maxima indicated with a yellow plus symbol.
The cyan contours enclose those local maxima that were undetected, while the red diamonds represent locations of local
maxima without a contour. The yellow dots represent locations of local minima. The left and right panels correspond to UDs identified
using the regular MLT and modified MLT technique, respectively. See the text for an explanation of the black dashed box.}
\label{fig03a}
\end{figure*}

\begin{figure*}[!ht] 
\centerline{
\hspace{40pt}
\includegraphics[angle = 90,width=0.55\textwidth]{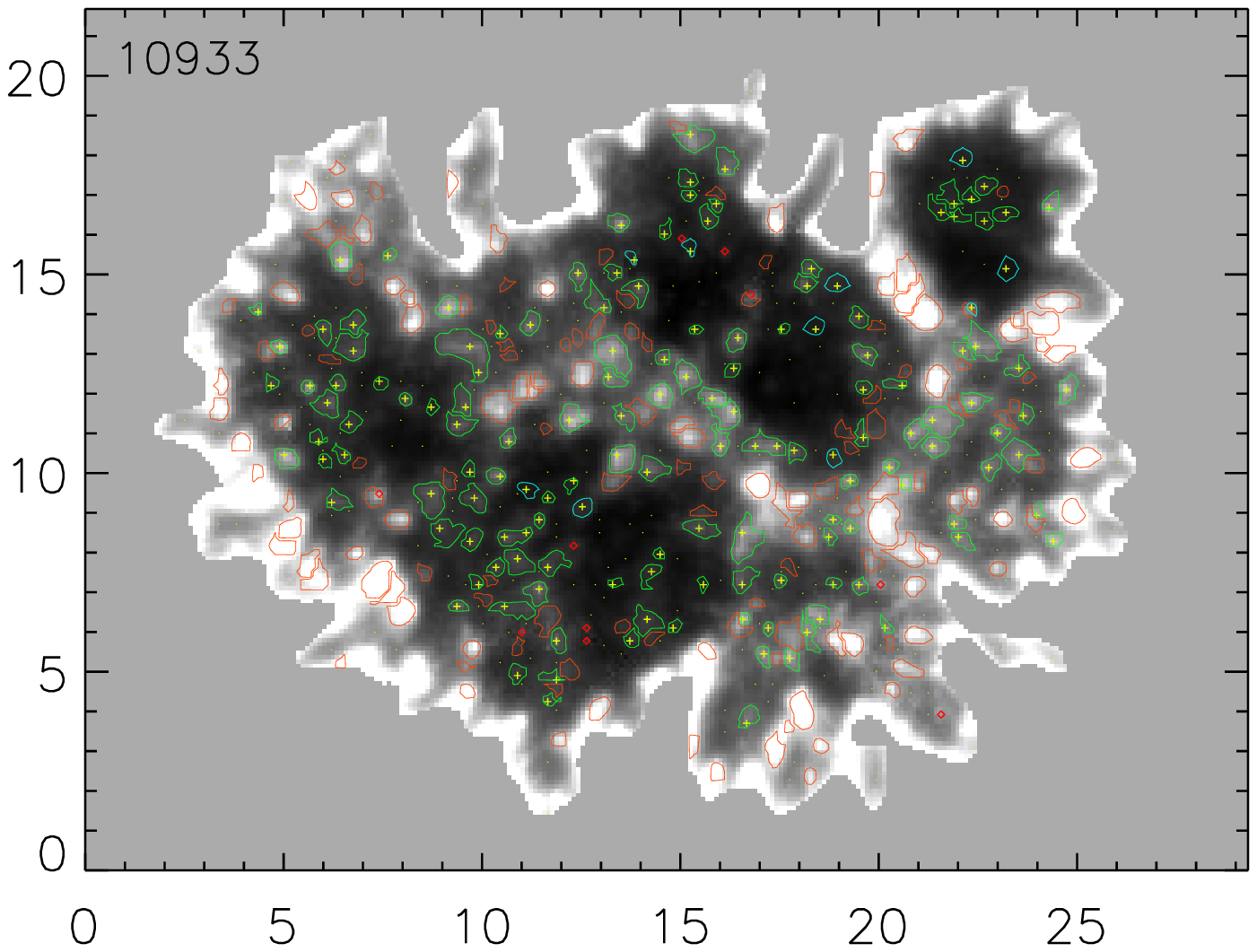}
\hspace{-50pt}
\includegraphics[angle = 90,width=0.55\textwidth]{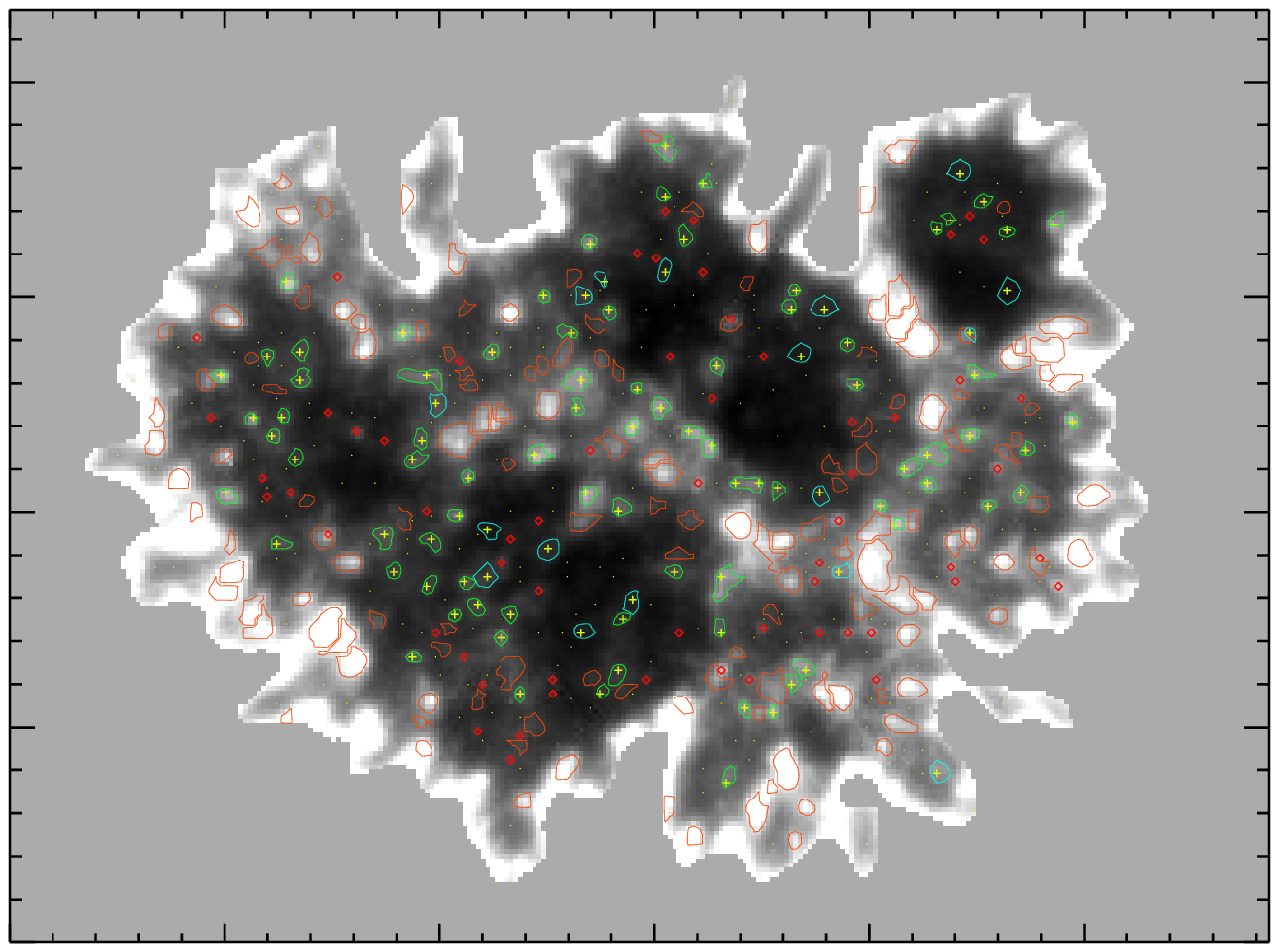}
}
\vspace{5pt}
\caption{Same as Fig.~\ref{fig03a} but for AR 10933.}
\label{fig03b}
\end{figure*}

\begin{figure*}[!ht] 
\centerline{
\hspace{145pt}
\includegraphics[angle = 90,width=0.84\textwidth]{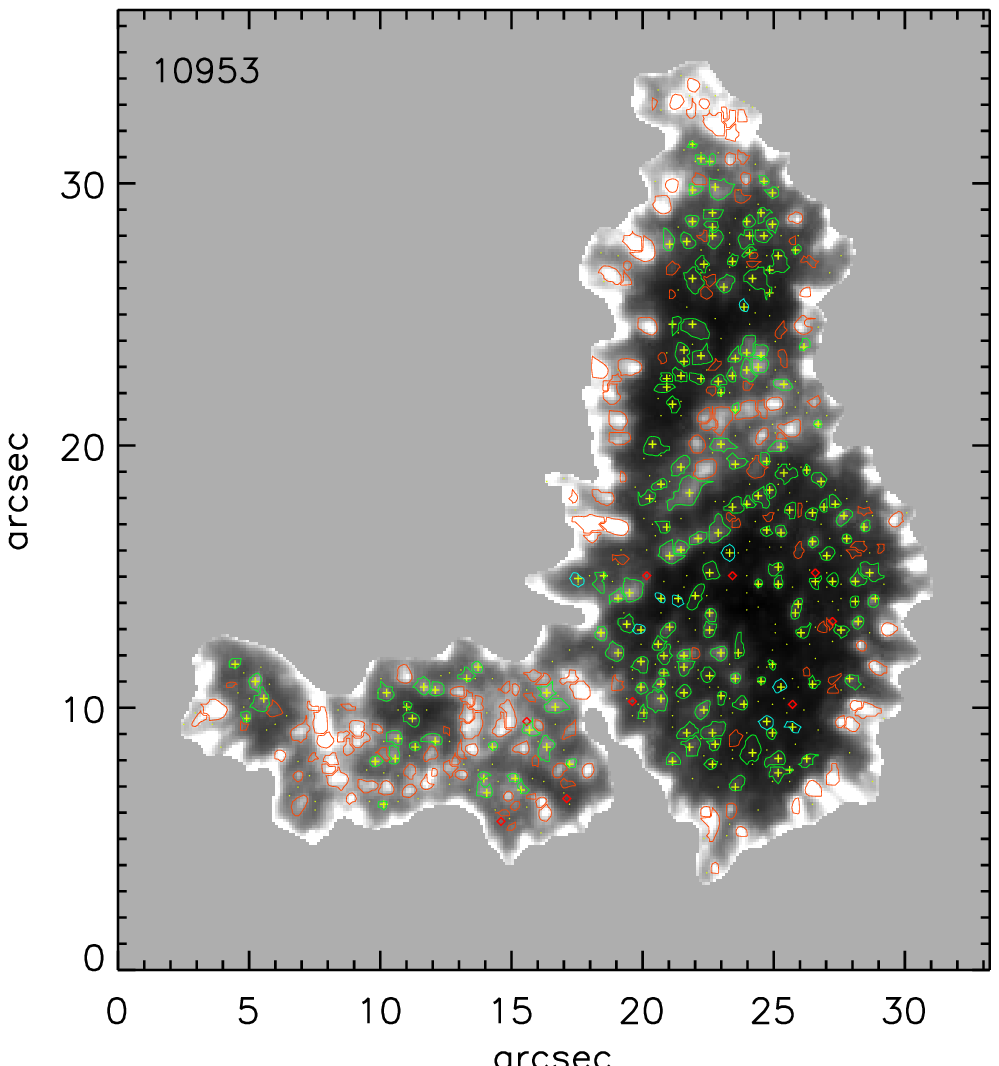}
\hspace{-190pt}
\includegraphics[angle = 90,width=0.84\textwidth]{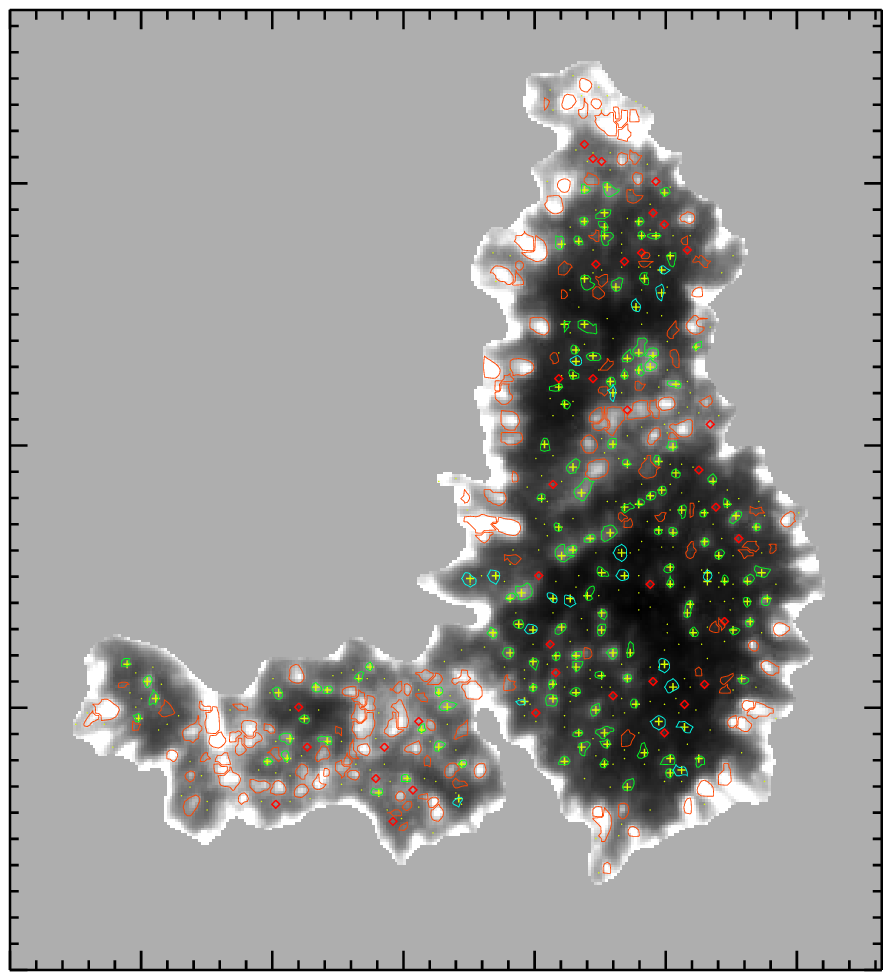}
}
\vspace{5pt}
\caption{Same as Fig.~\ref{fig03b} but for AR 10953.}
\label{fig03c}
\end{figure*}

\begin{figure*}[!ht] 
\centerline{
\hspace{-20pt}
\includegraphics[angle = 90,width=0.5\textwidth]{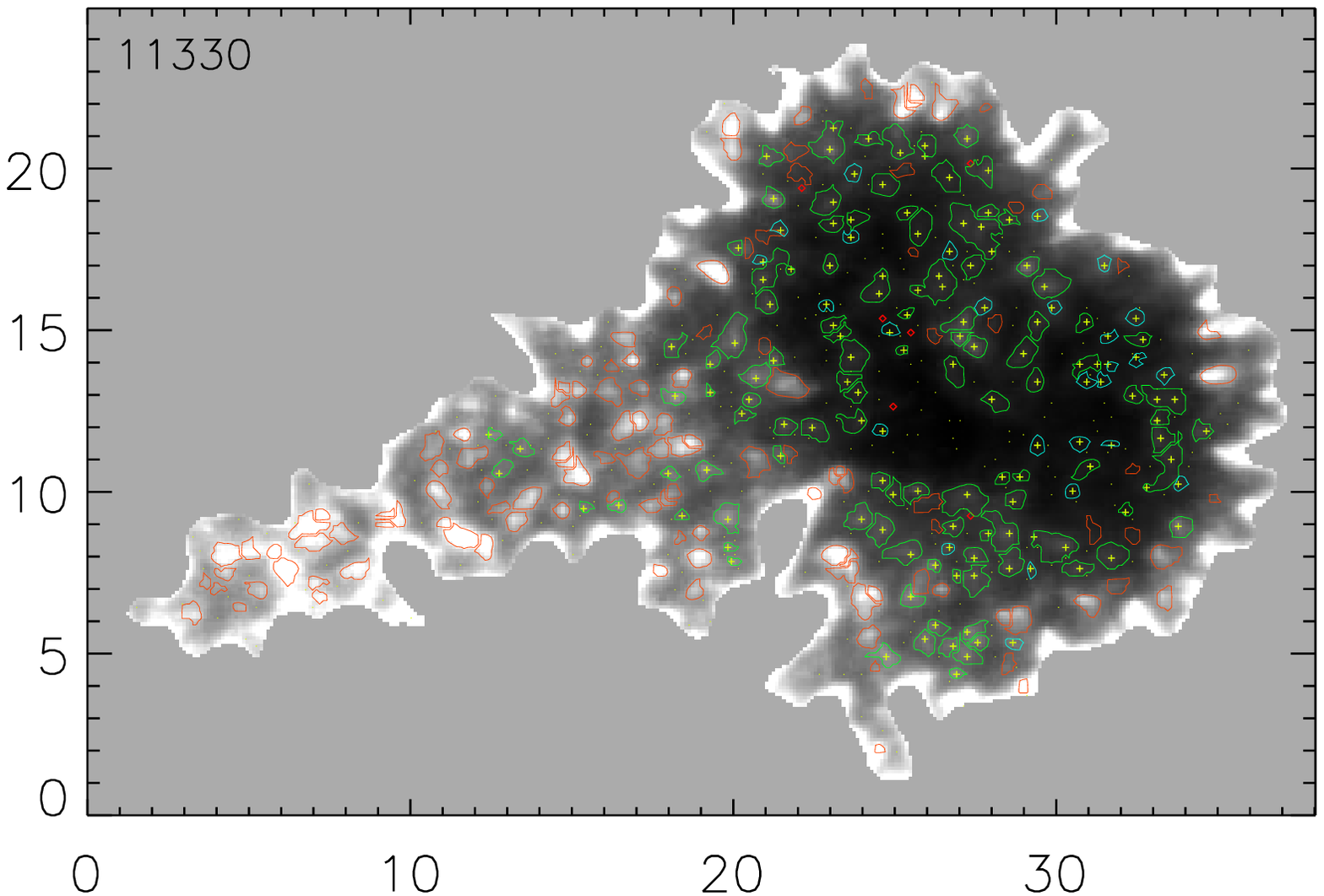}
\hspace{-28pt}
\includegraphics[angle = 90,width=0.5\textwidth]{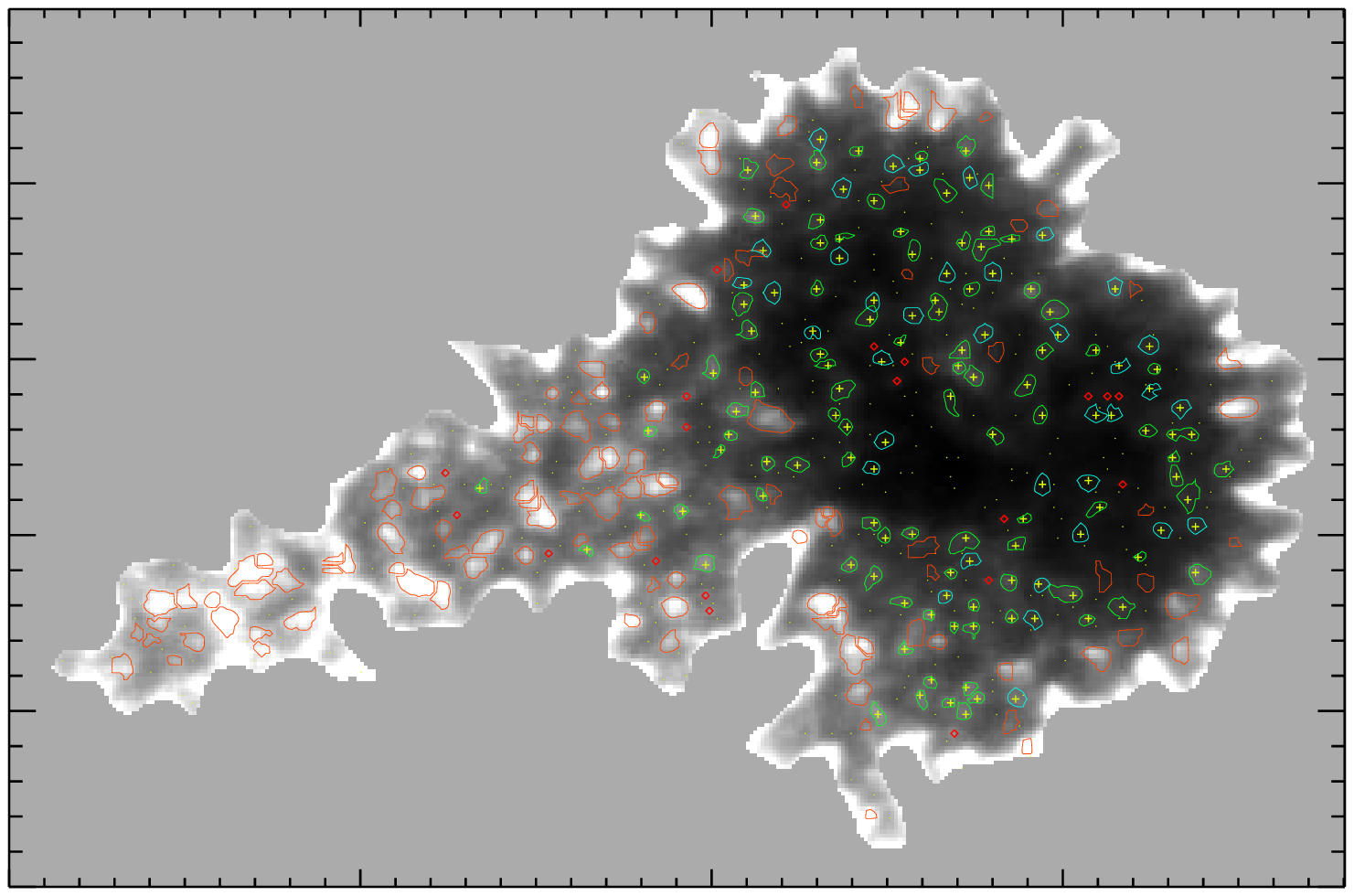}
}
\vspace{5pt}
\caption{Same as Fig.~\ref{fig03c} but for AR 11330.}
\label{fig03d}
\end{figure*}

\begin{figure*}[!ht] 
\centerline{
\includegraphics[angle = 0,width=0.5\textwidth]{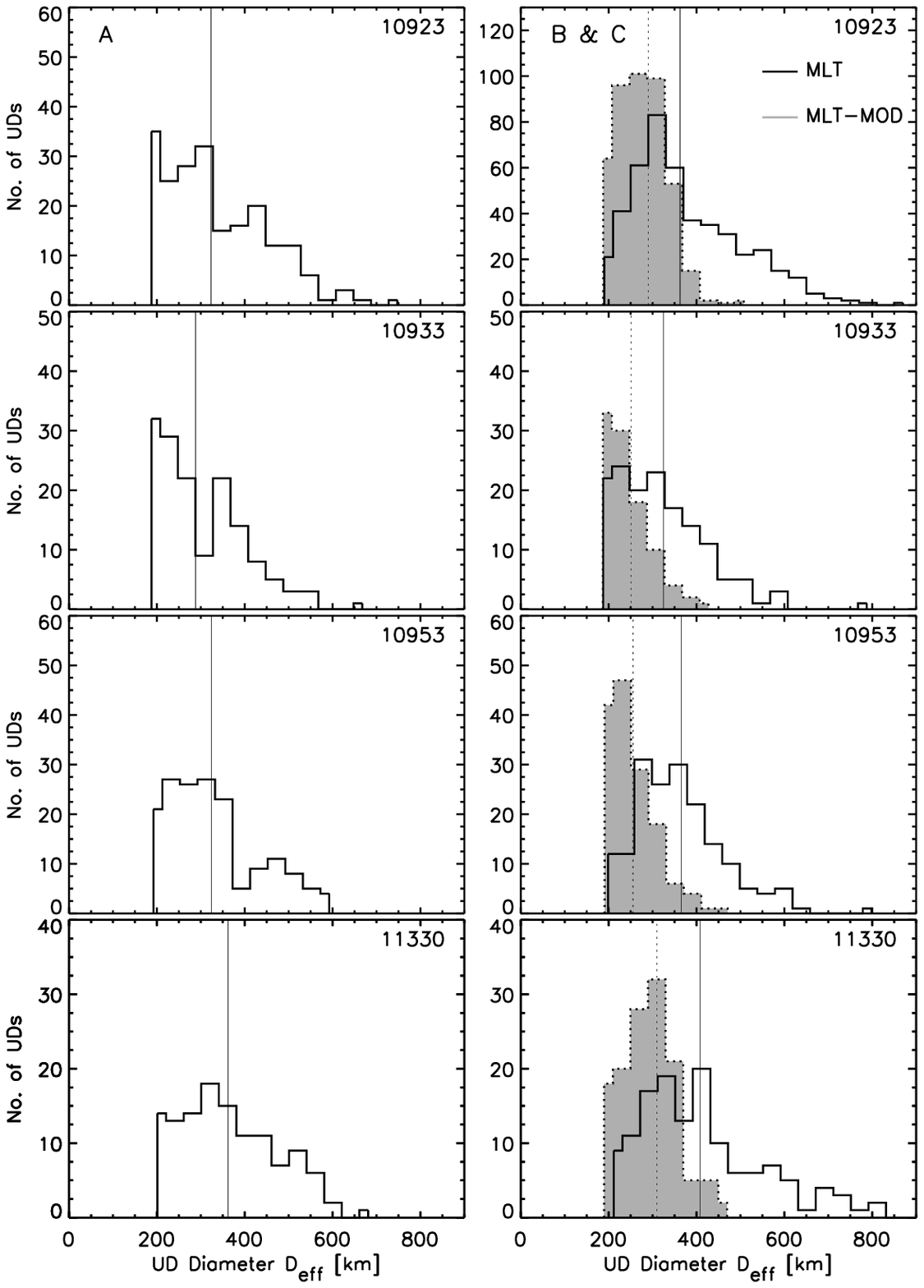}
\includegraphics[angle = 0,width=0.5\textwidth]{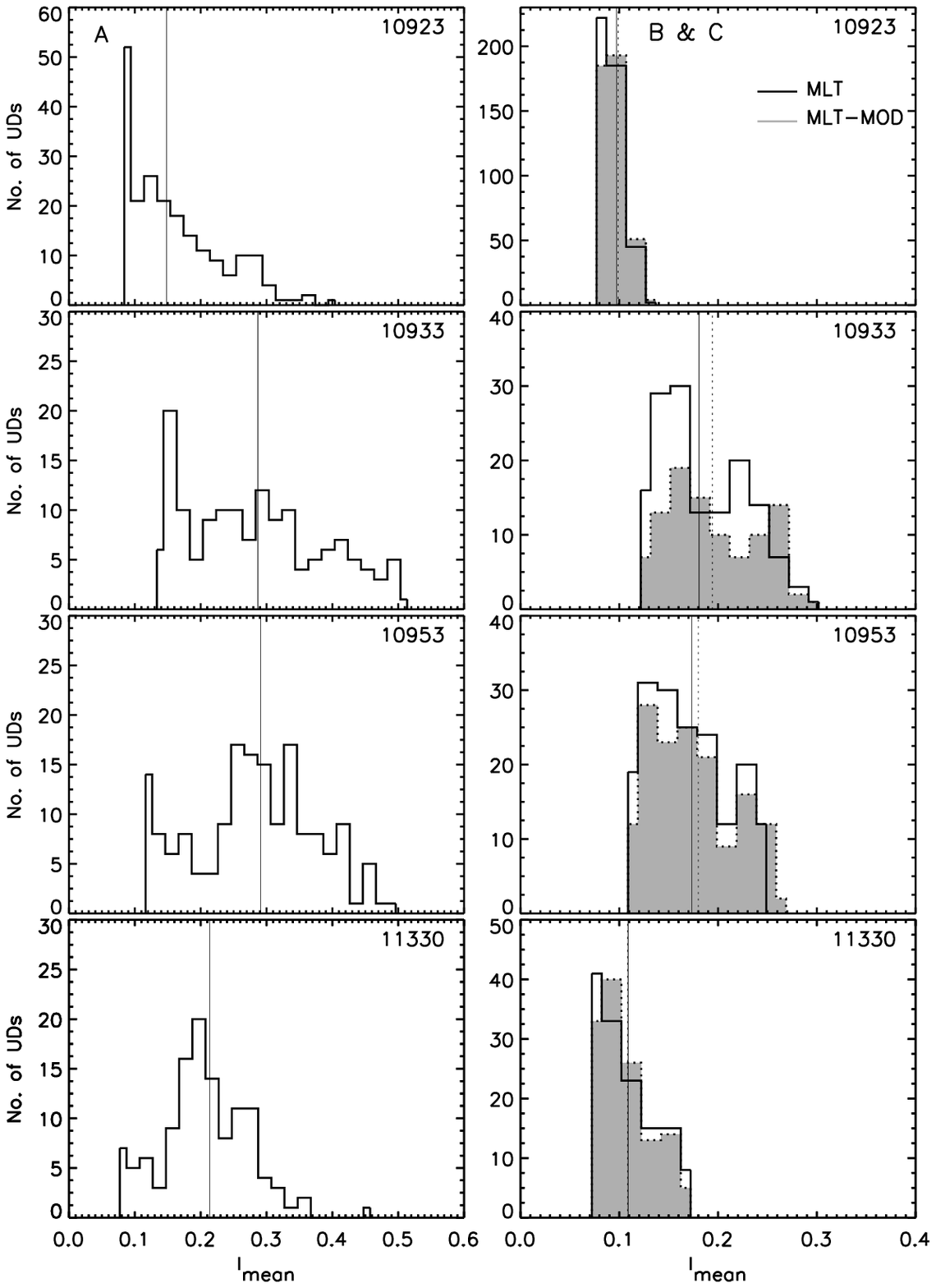}
}
\vspace{5pt}
\caption{Distribution of effective diameter (columns 1 and 2) and mean intensity (columns 3 and 4) of UDs. The
left column corresponds to the red contours, labeled ``A'', while the right column corresponds to the green and cyan contours, labeled ``B'' 
and ``C'', respectively. The black open and gray shaded histograms correspond to the quantities derived using the regular and modified MLT scheme, 
respectively. A bin size of 40\,km and 0.02 was used in the histograms of the effective diameter and mean intensity, respectively.}
\label{fig05}
\end{figure*}

\section{Results}
\label{results}

\subsection{Spatial Distribution of UDs}
\label{spatial}
Figures~\ref{fig03a}--\ref{fig03d} show the identified UDs in the different sunspots using the regular (left) and modified (right) 
MLT scheme. The red contours are common to both schemes and primarily trace brighter UDs that lie near the umbra-penumbra boundary 
as well as in the extended structures such as light bridges. This is due to the criteria set for the contours labeled ``A'', which 
do not enclose any points of maximum intensity in the local neighborhood. The UDs enclosed by the red contours account for nearly 
half the total number in ARs 10933, 10953, and 11330 using both detection schemes, while it is around 30\% for AR 10923. 
The number of UDs detected by the two schemes differ nearly entirely in the green contours and only marginally in the cyan contours, 
which correspond to contours labeled ``B'' and ``C'', respectively, with the regular
scheme identifying more structures in the green contours than the modified scheme. The green and cyan contours enclose UDs that are 
located in the relatively dark or diffuse regions of the umbra, often in the spaces between brighter UDs or on the flanks of light 
bridges. In addition, the figures also show that the green contours in the regular scheme tend to be larger than their counterparts 
in the modified scheme due to the threshold value corresponding to 50\% of the sum of the maximum and background intensity. The 
difference in the number of UDs detected by the two schemes with respect to the regular MLT scheme is less than 10\% for ARs 
10923, 10953, and 11330, while it is around 16\% for AR 10923. Some contours, specifically the ``A''-labeled ones, can appear stacked
in succession, as shown in the black dashed box in Fig.~\ref{fig03a}. An inset of this region shown on the left indicates that it is due
to multiple structures that appear alongside each other which is an indication that the MLT detection is not spurious. 
In the following sections, we will refer either to the color or labels of the contours analogously.    

\subsection{Physical Characteristics of UDs}
\label{properties}
Figure~\ref{fig05} shows the histograms for the effective diameter ($D_{\textrm{\tiny{eff}}}$) and mean intensity ($I_{\textrm{\tiny{mean}}}$) 
for the contours labeled ``A'' (left) and those labeled ``B'' and ``C'' (right). The median value of $D_{\textrm{\tiny{eff}}}$ varies between 290--360\,km for 
the contours labeled ``A'', with the majority of UDs in all ARs having a value less than 400\,km. However, for UDs enclosed by contours ``B'' and ``C'', 
there is a clear distinction in the distribution as retrieved by the regular (open; solid) and modified MLT scheme (dark shaded; dotted). 
The median values obtained by the regular scheme are on average 70--90\,km greater than those obtained by the modified scheme, and the 
histograms tend to extend beyond 400\,km, but the number of UDs with $D_{\textrm{\tiny{eff}}}>600$\,km is less than 10, as seen in ARs 10923 and 11330.
The histogram of $D_{\textrm{\tiny{eff}}}$ for contours ``B'' and ``C'' from the modified scheme is relatively narrow and confined to a range of 
about 200--400\,km, with median values ranging between 250 and 310\,km for the four sunspots. On the other hand, the median values of 
$D_{\textrm{\tiny{eff}}}$ for the contours labeled ``B'' and ``C'' vary from 320 to 410\,km for the regular scheme and are on average 40\,km larger than UDs enclosed by 
contours labeled ``A''. If one considers all contours combined, then the median value of $D_{\textrm{\tiny{eff}}}$ from the regular approach 
is around 50\,km larger than those retrieved by the modified approach. 

\begin{figure}[!ht] 
\centerline{
\includegraphics[angle = 0,width=\columnwidth]{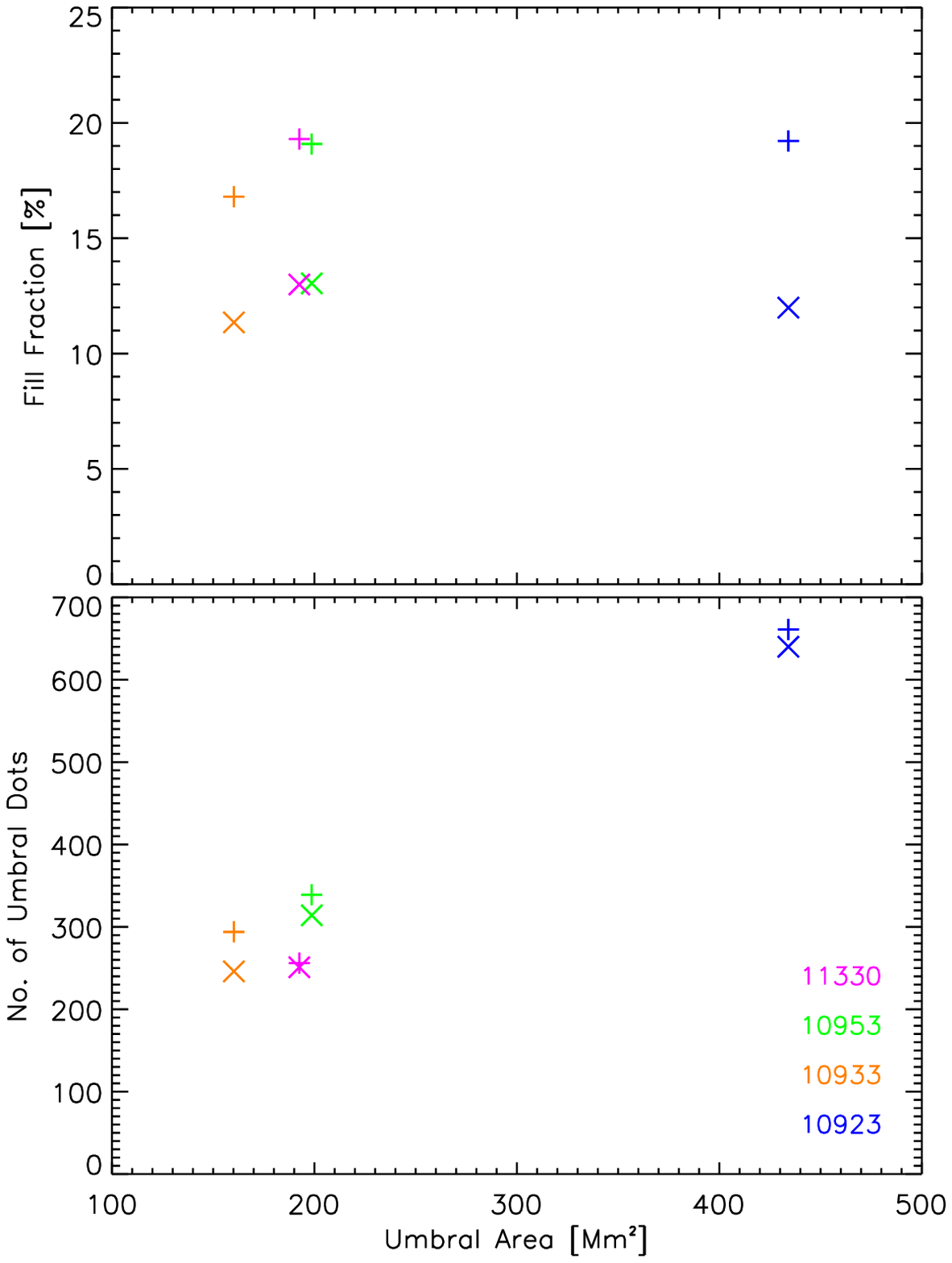}
}
\vspace{-10pt}
\caption{Variation of the UD fill fraction and number as a function of umbral area. The plus and cross symbols refer to the quantities
derived from the regular and modified MLT scheme, respectively.}
\label{fig06}
\end{figure}

The histograms of the mean intensity ($I_{\textrm{\tiny{mean}}}$) of UDs enclosed by the red contours (column 3) tend to vary with the 
sunspots, with a central peak seen in ARs 10953 and 11330, while AR 10923 and 10933 show a large population of fainter UDs. The range of 
intensities is also different for the four spots, with values extending up to 0.45 for ARs 10933 and 10953, while it is confined to about 
0.3 for ARs 10923 and 11330. The median values range from 0.15 in AR 10923 to 0.29 in ARs 10933 and 10953. On the other hand, the histograms 
of the mean intensity of UDs enclosed by the green and cyan contours (column 4) for both MLT schemes are nearly identical and tend to be narrow, 
as seen in ARs 10923 and 11330, or extended, as in ARs 10933 and 10953. The median values of the mean intensity for UDs within contours labeled 
``B'' and ``C'' vary from 0.1--0.19, but are nearly the same for both MLT schemes, with the exception of AR 10933, where only the histogram 
amplitudes differ between the two approaches.

The quantity $I_{\textrm{\tiny{max}}}/I_{\textrm{\tiny{bck}}}$ exhibits an extended histogram (not shown) ranging between 1.02 and 2.5 with 
median values ranging between 1.42 and 1.62 for the UDs enclosed by the contours labeled ``A''. Unlike $D_{\textrm{\tiny{eff}}}$, the histograms of
$I_{\textrm{\tiny{max}}}/I_{\textrm{\tiny{bck}}}$ for UDs enclosed by contours labeled ``B'' and ``C'', are nearly identical for both the regular 
and modified MLT schemes, with only AR 10933 showing a difference in values around 1.2--1.5 but with a nearly identical shape. The differences 
in the median values for both schemes vary between 1.2 and 1.43, and the difference is less than 10\%.

Figure~\ref{fig06} shows the variation of the fill fraction and number of UDs with the 
area of the umbral core. The fill fraction is obtained as the ratio of the total area enclosed by the UD contours to the umbral area 
expressed as a percentage. The plus symbols correspond to the regular MLT scheme where the fill fraction varies from 17 to 19\%, while the 
cross symbols correspond to the modified MLT scheme where the values range from 12 to 13\%. The fill fraction obtained from both schemes 
is thus independent of the umbral area. The number of UDs on the other hand exhibits a very strong correlation with the umbral area, 
with a linear correlation coefficient of 0.98 and 0.99 for the regular and modified schemes, respectively.

\begin{figure*}[!ht] 
\centerline{
\includegraphics[angle = 0,width=0.5\textwidth]{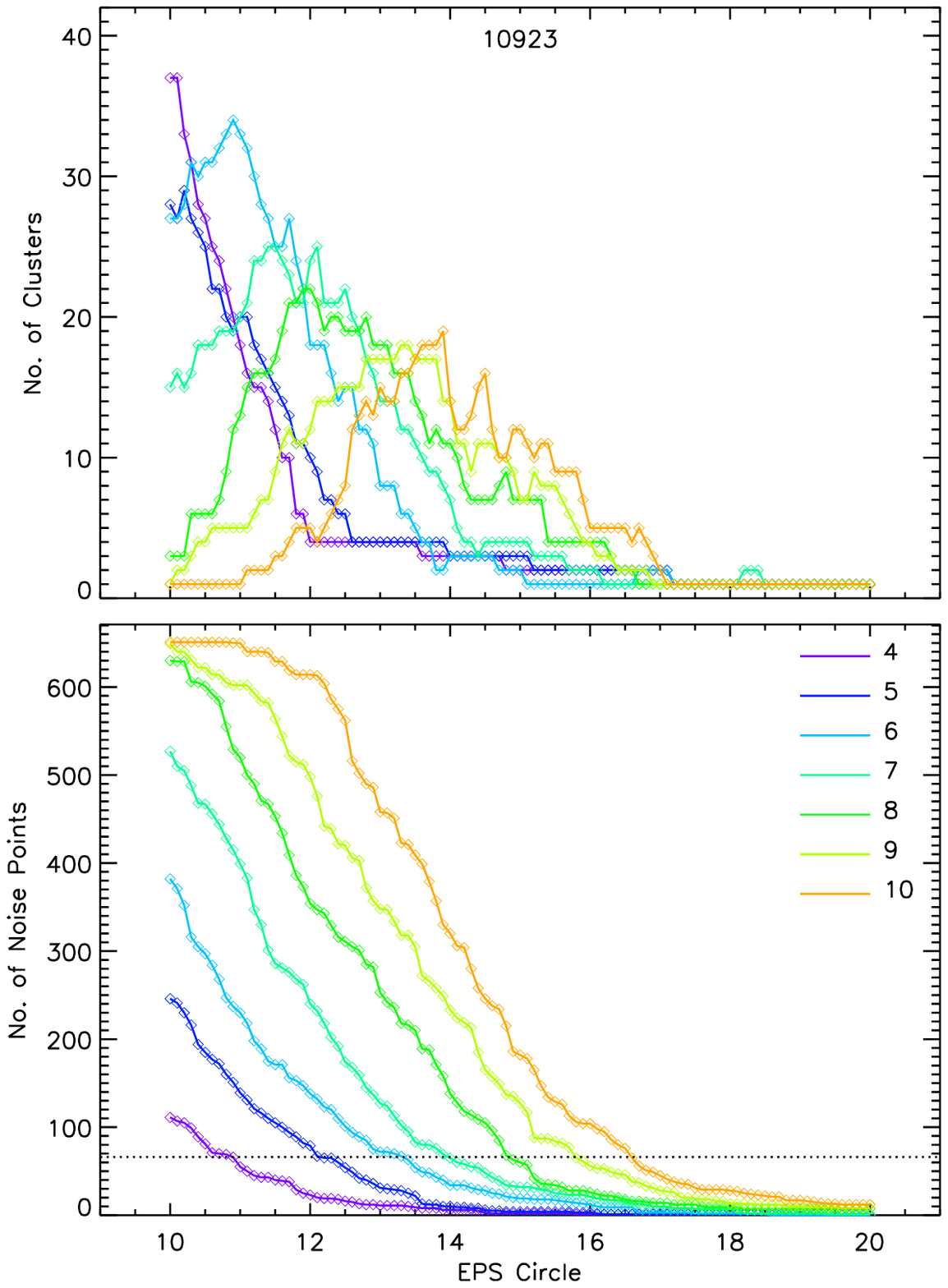}
\includegraphics[angle = 0,width=0.5\textwidth]{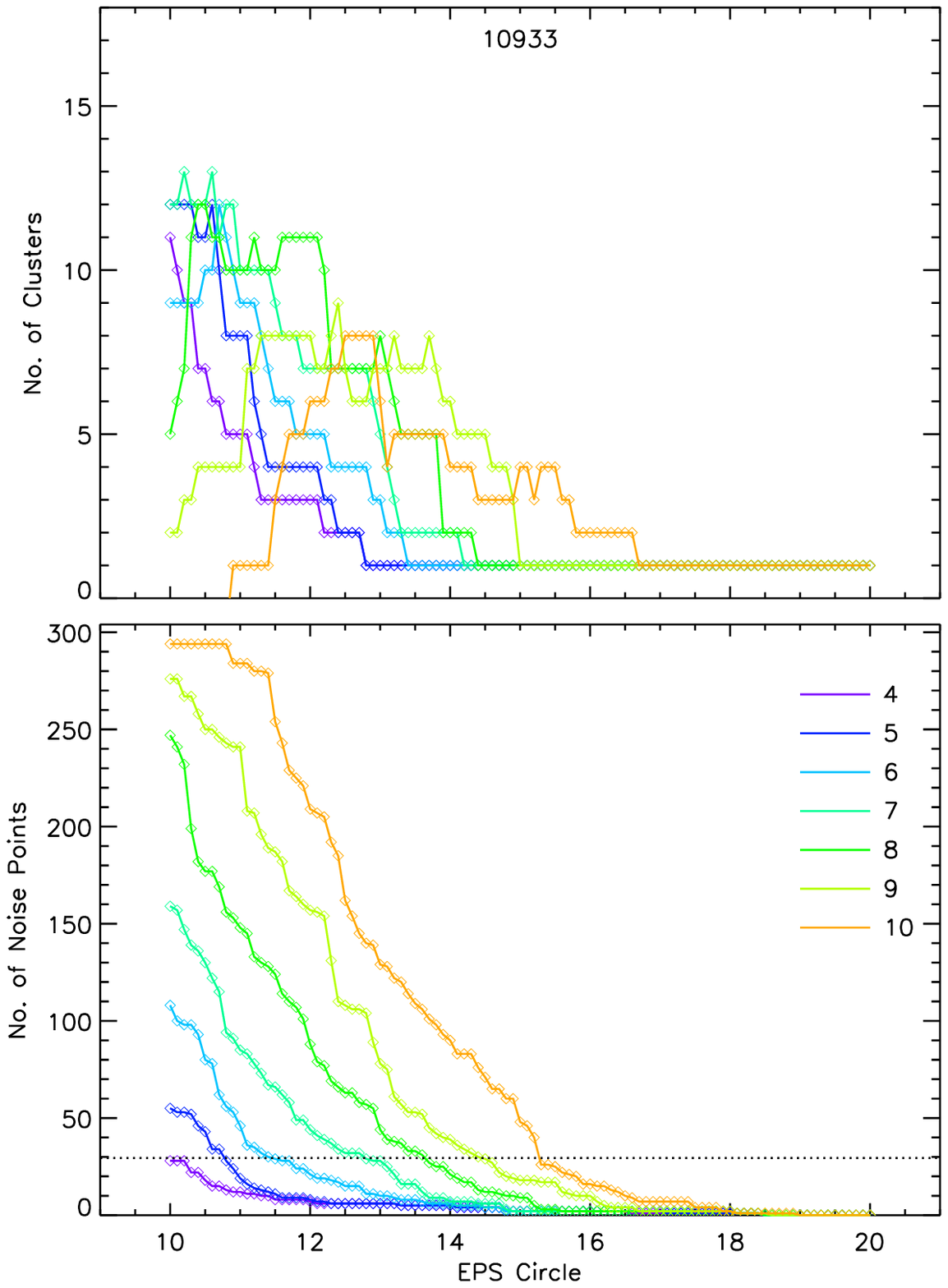}
}
\vspace{-10pt}
\caption{Variation of the number of clusters (top) and noise points (bottom) with the eps circle ($\epsilon$). The different 
colors correspond to the min. points ($M$). The left and right columns correspond to AR 10923 and 10933, respectively. The 
dotted line in the bottom panels represents 10\% of the total data points (UD centroids), which is five units less than the maximum value on the $y$-axis.}
\label{fig07}
\end{figure*}

\begin{figure*}[!ht] 
\centerline{
\includegraphics[angle = 0,width=0.5\textwidth]{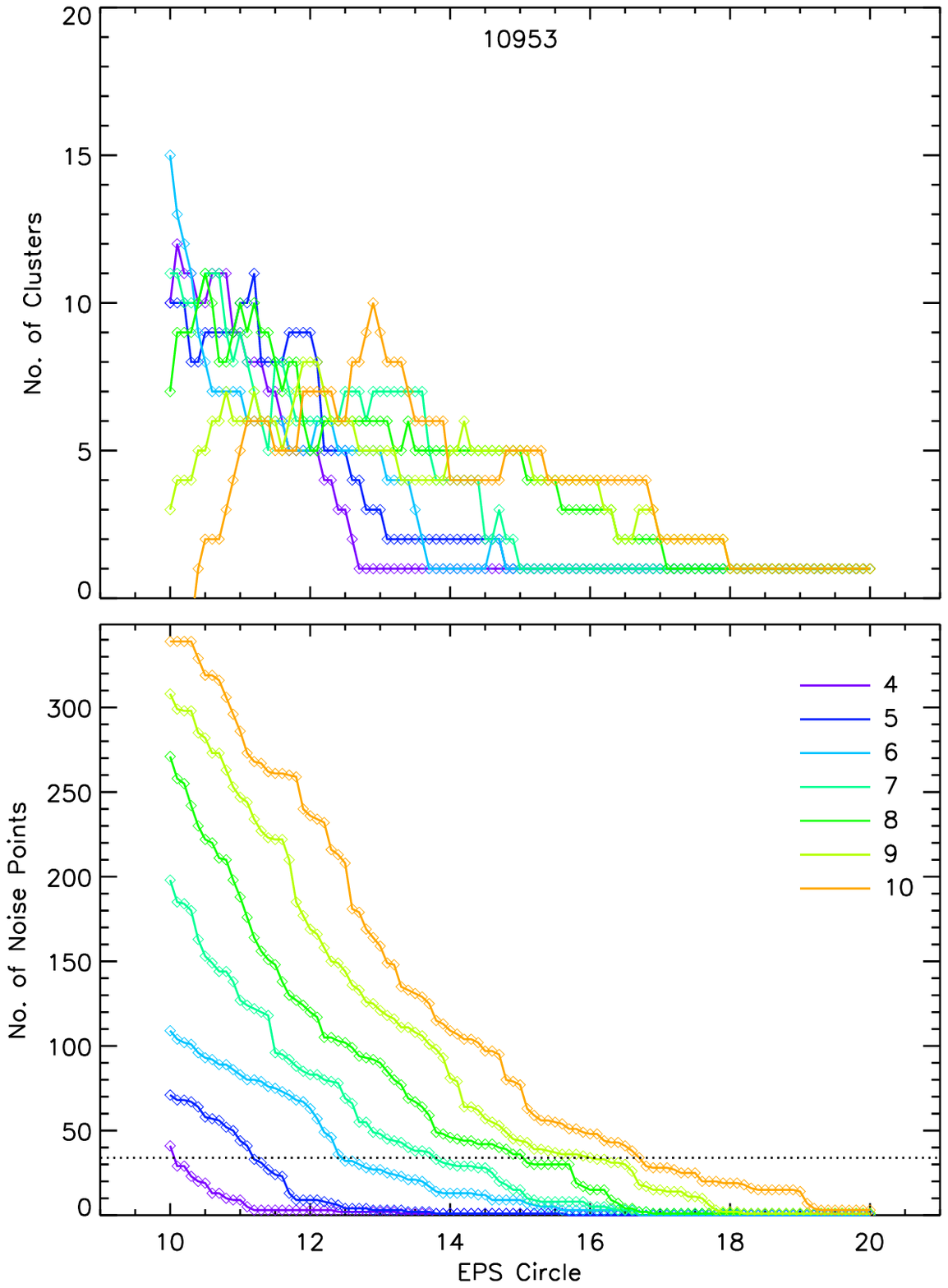}
\includegraphics[angle = 0,width=0.5\textwidth]{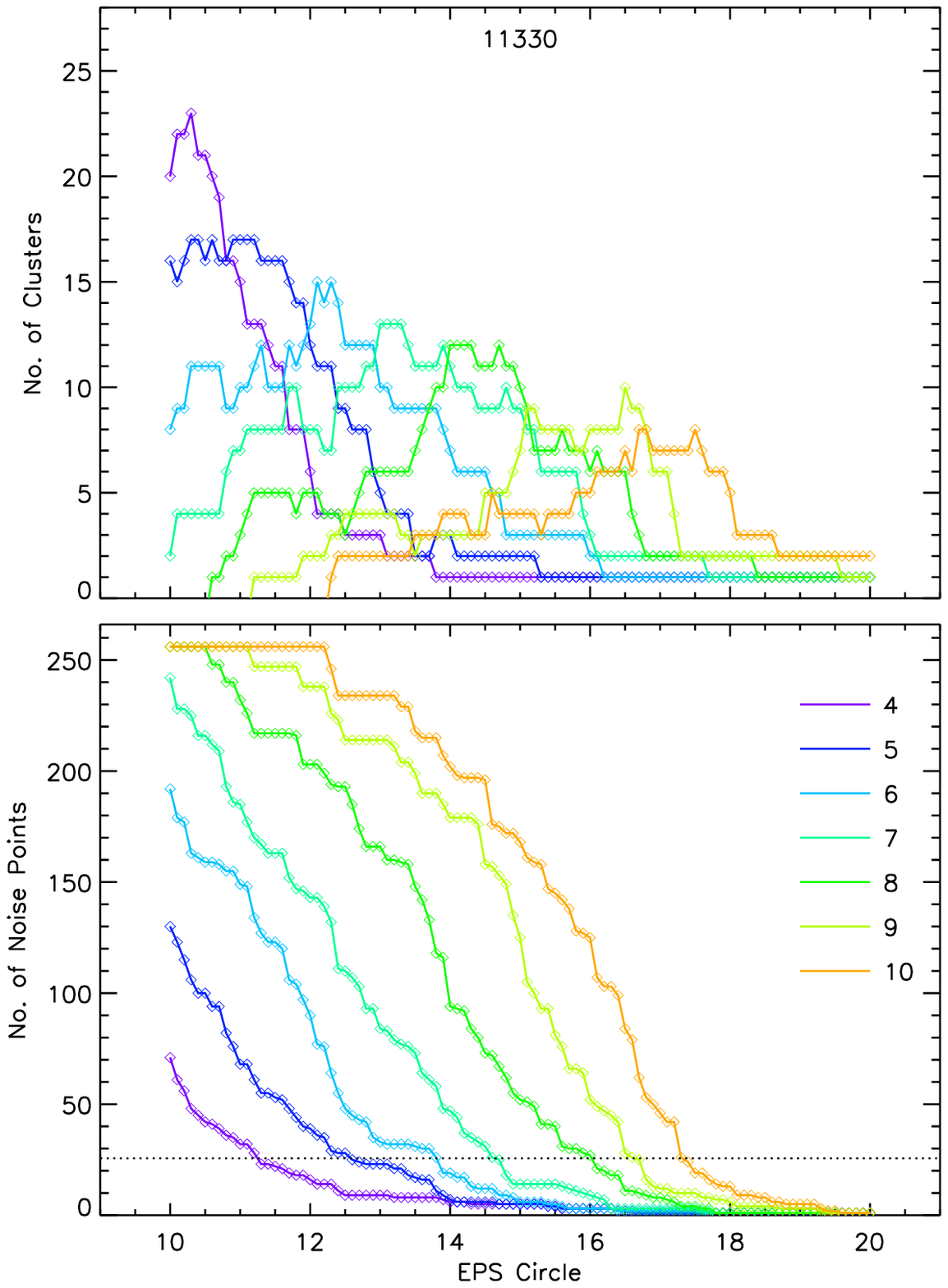}
}
\vspace{-10pt}
\caption{Same as Fig.~\ref{fig07} but for AR 10953 (left) and 11330 (right).}
\label{fig08}
\end{figure*}

\begin{figure*}[!ht] 
\centerline{
\hspace{145pt}
\includegraphics[angle = 90,width=0.8\textwidth]{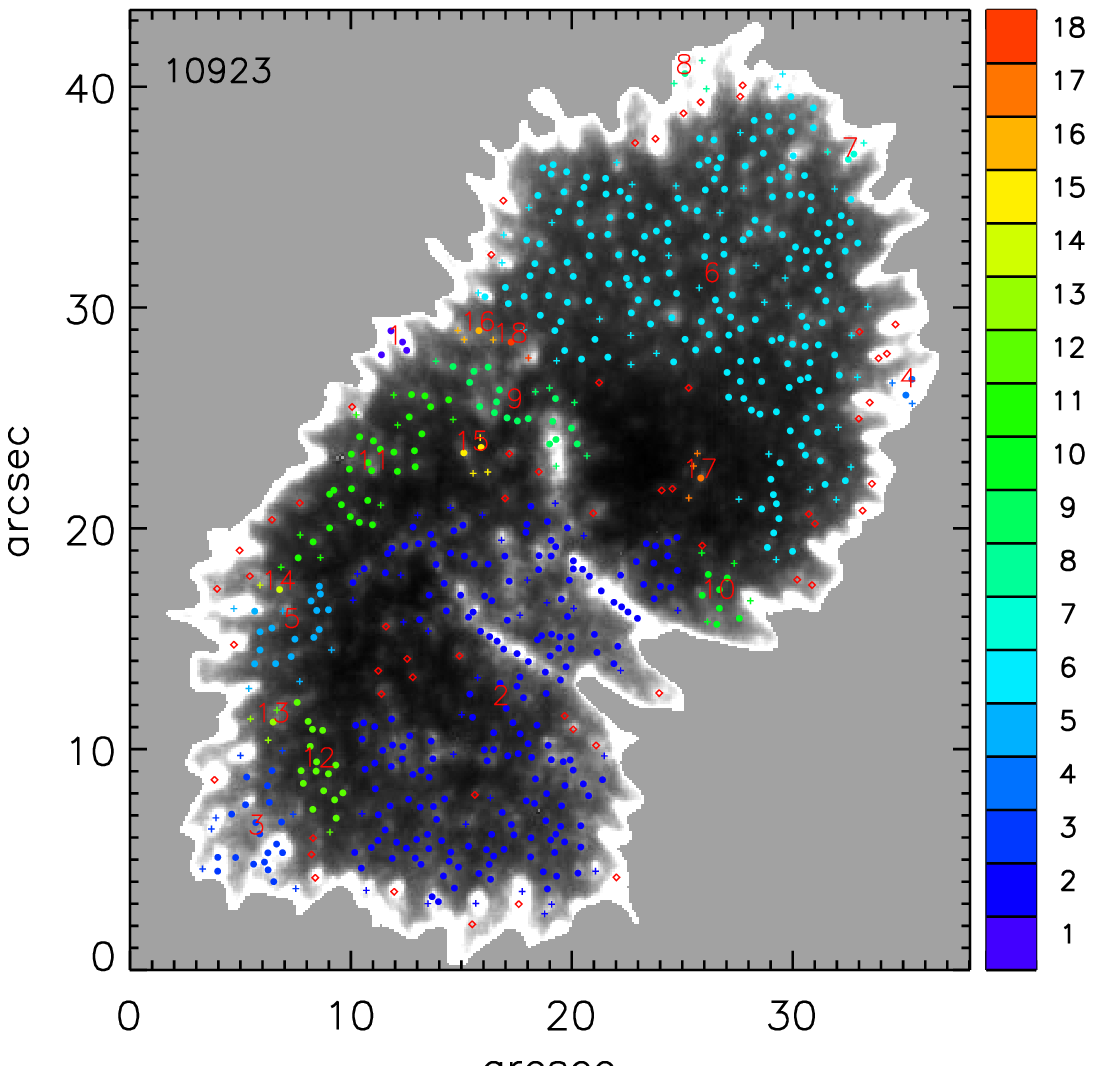}
\hspace{-165pt}
\includegraphics[angle = 90,width=0.8\textwidth]{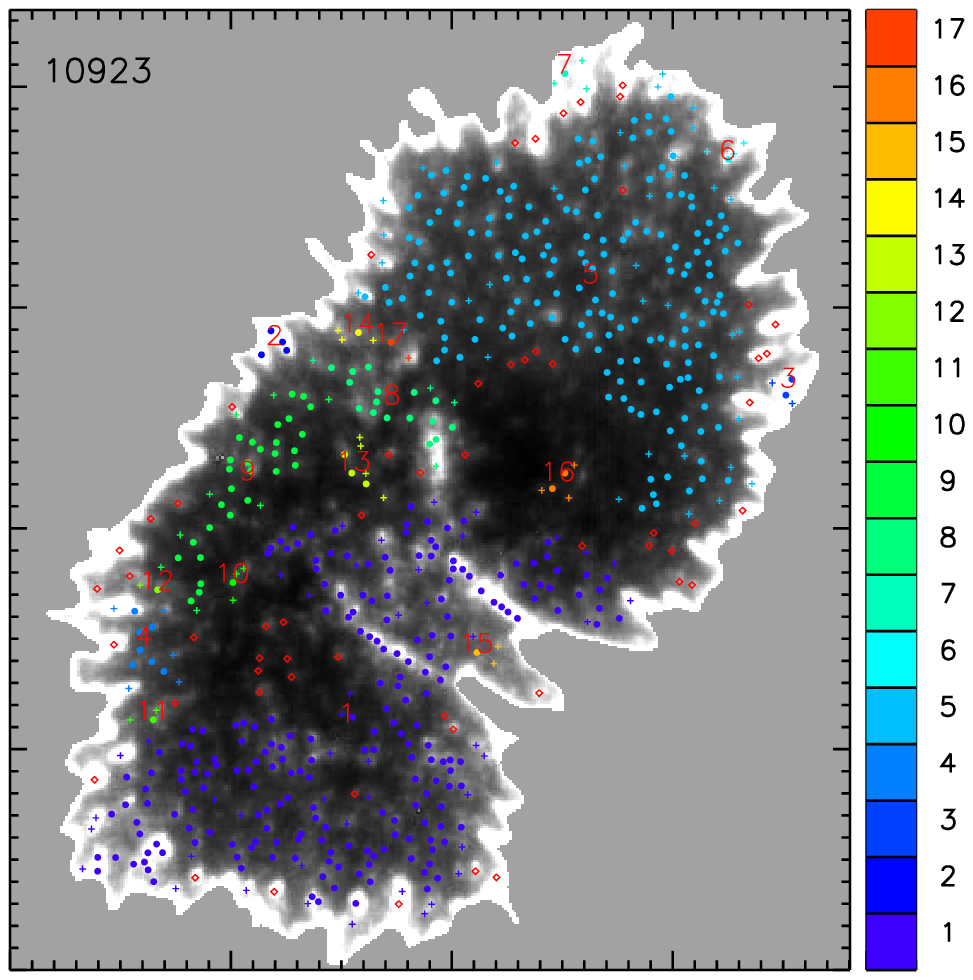}
}
\vspace{-5pt}
\centerline{
\hspace{30pt}
\includegraphics[angle = 90,width=0.53\textwidth]{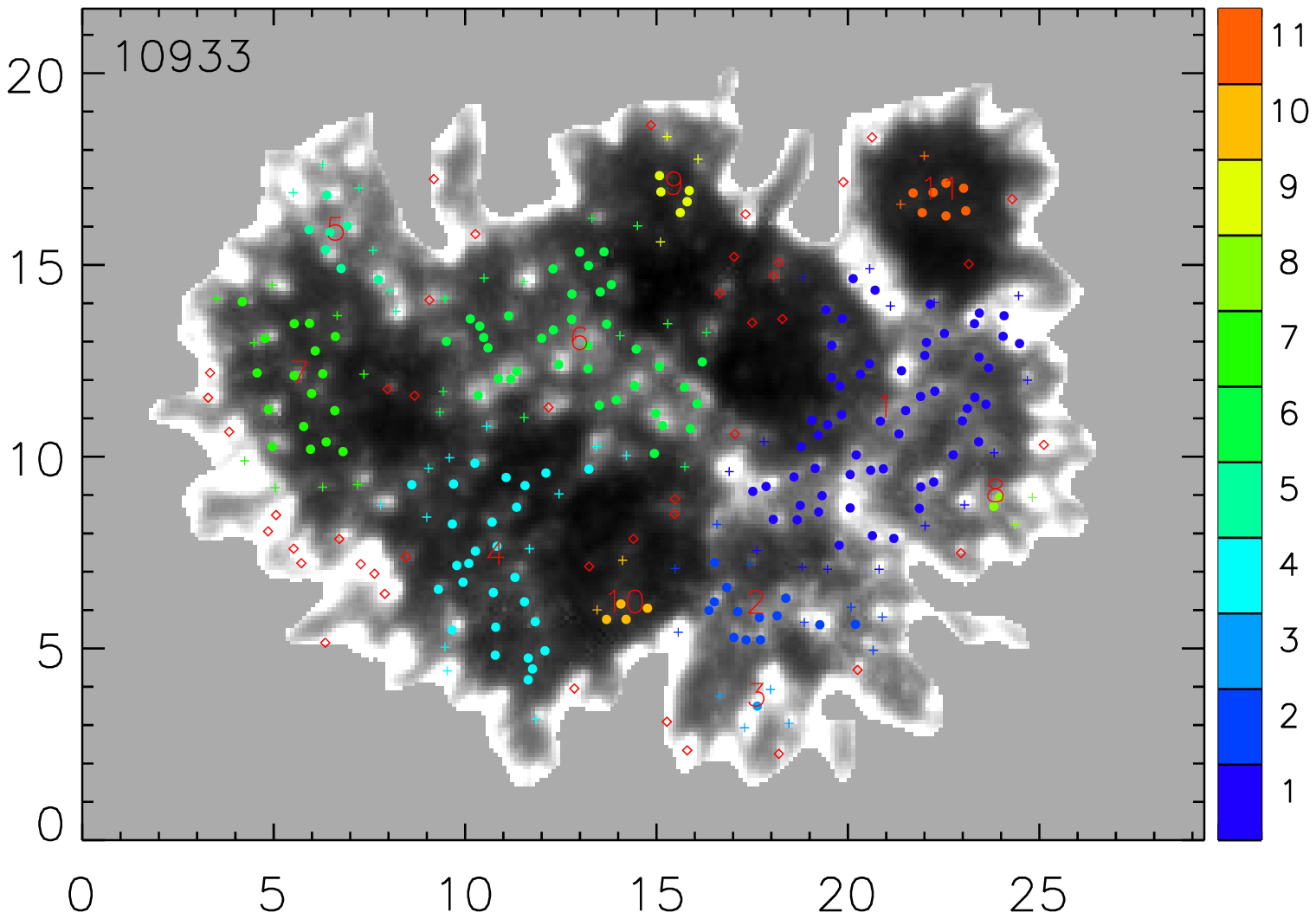}
\hspace{-35pt}
\includegraphics[angle = 90,width=0.53\textwidth]{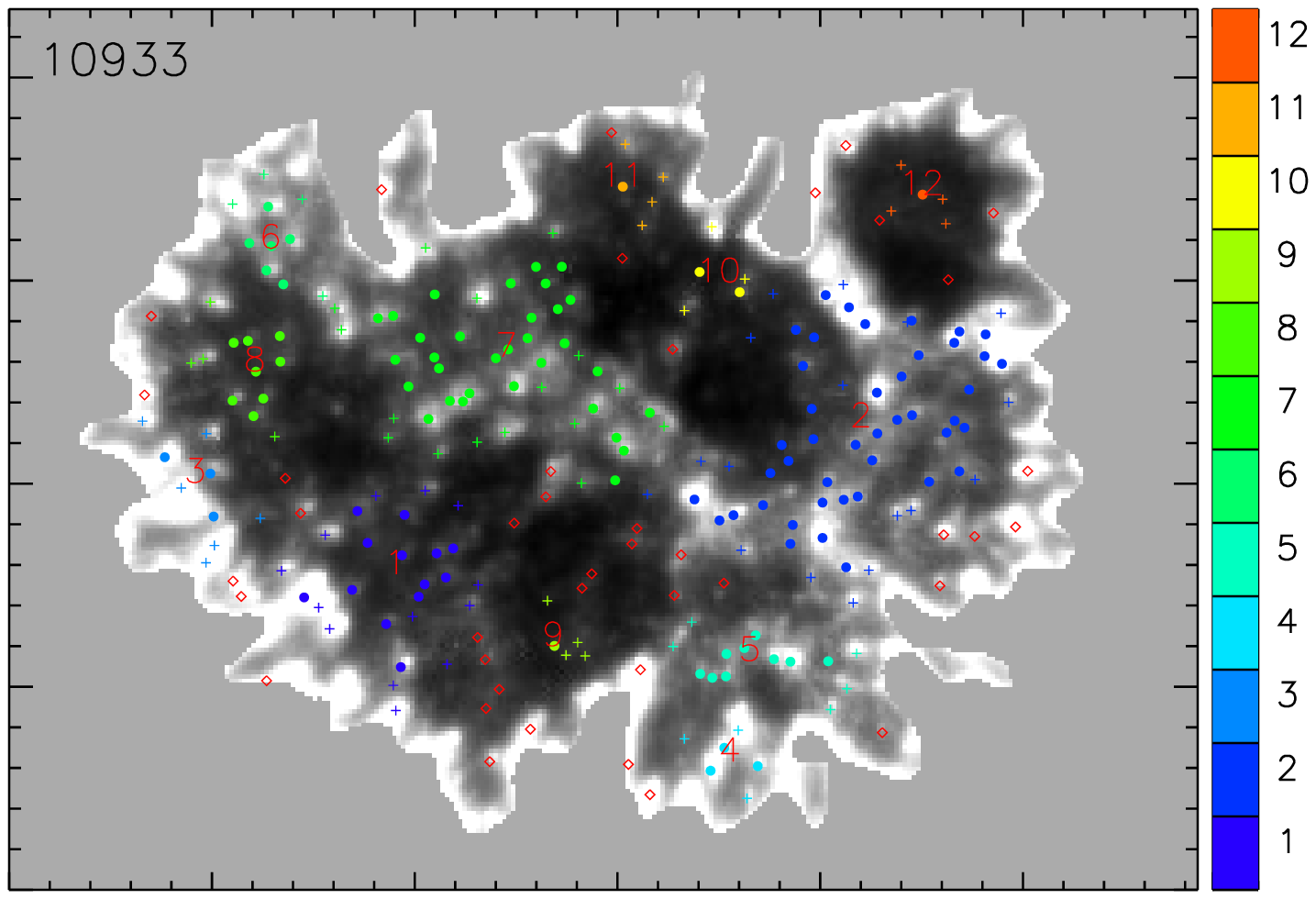}
}
\vspace{10pt}
\caption{Clustering of UDs derived from DBSCAN. The colored filled circles represent core points, 
as indicated in the legend on the right, while the plus symbols represent border points. The red diamonds
correspond to noise points. The top panels correspond to AR 10923, with the left and right panel showing the clustering with the regular 
and modified MLT scheme, respectively. Bottom panels: same as the top panels but for AR 10933.}
\label{fig09}
\end{figure*}

\begin{figure*}[!ht] 
\centerline{
\hspace{130pt}
\includegraphics[angle = 90,width=0.81\textwidth]{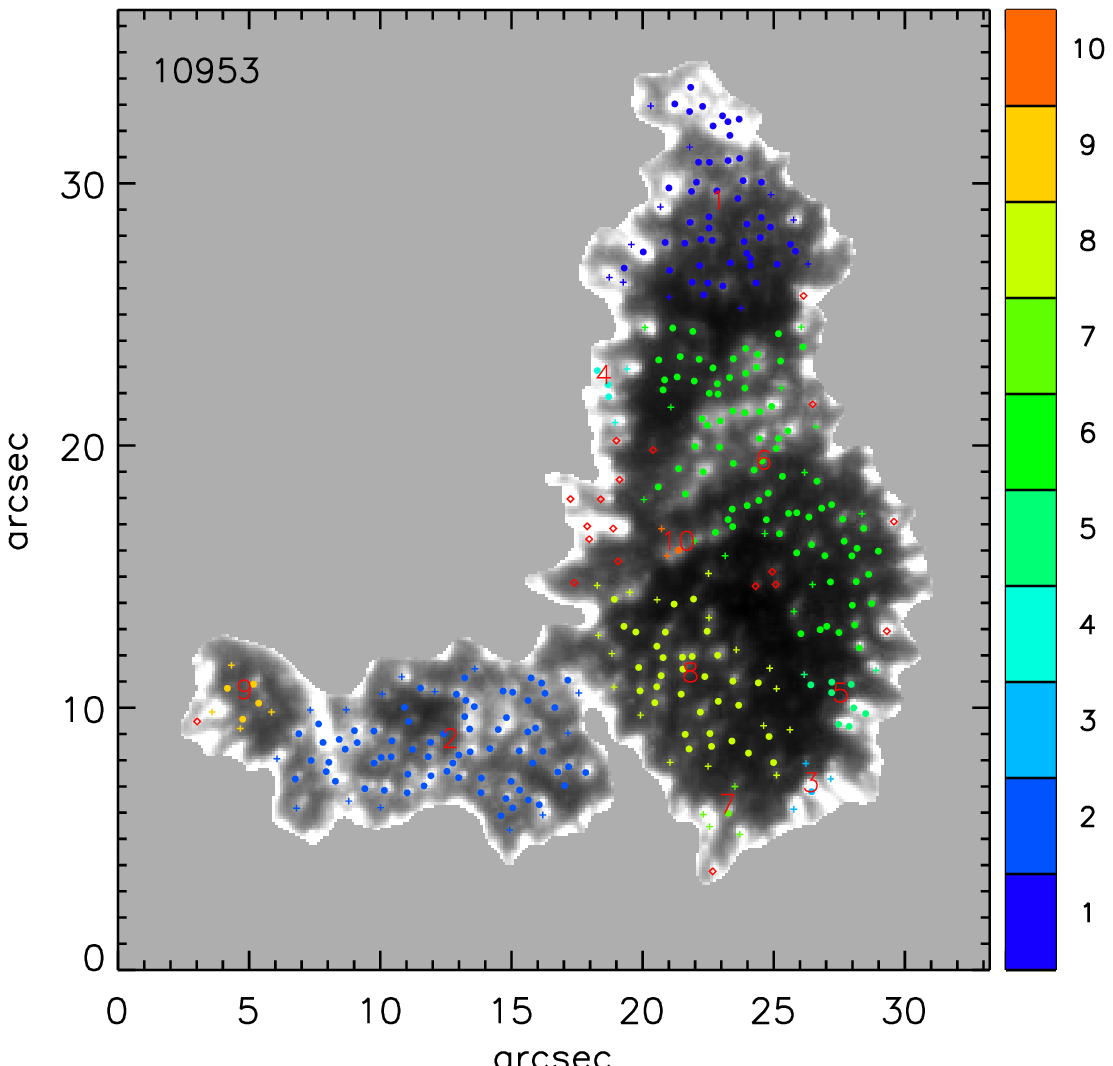}
\hspace{-160pt}
\includegraphics[angle = 90,width=0.81\textwidth]{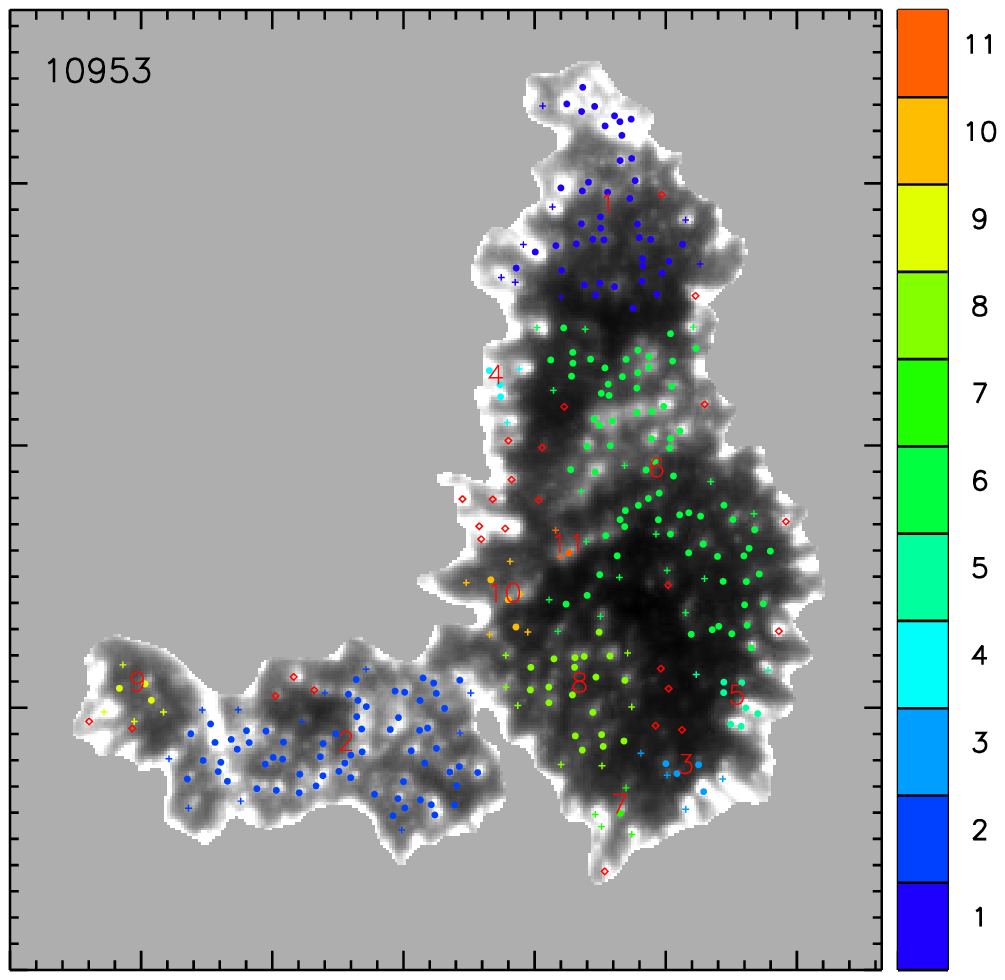}
}
\vspace{-5pt}
\centerline{
\hspace{0pt}
\includegraphics[angle = 90,width=0.5\textwidth]{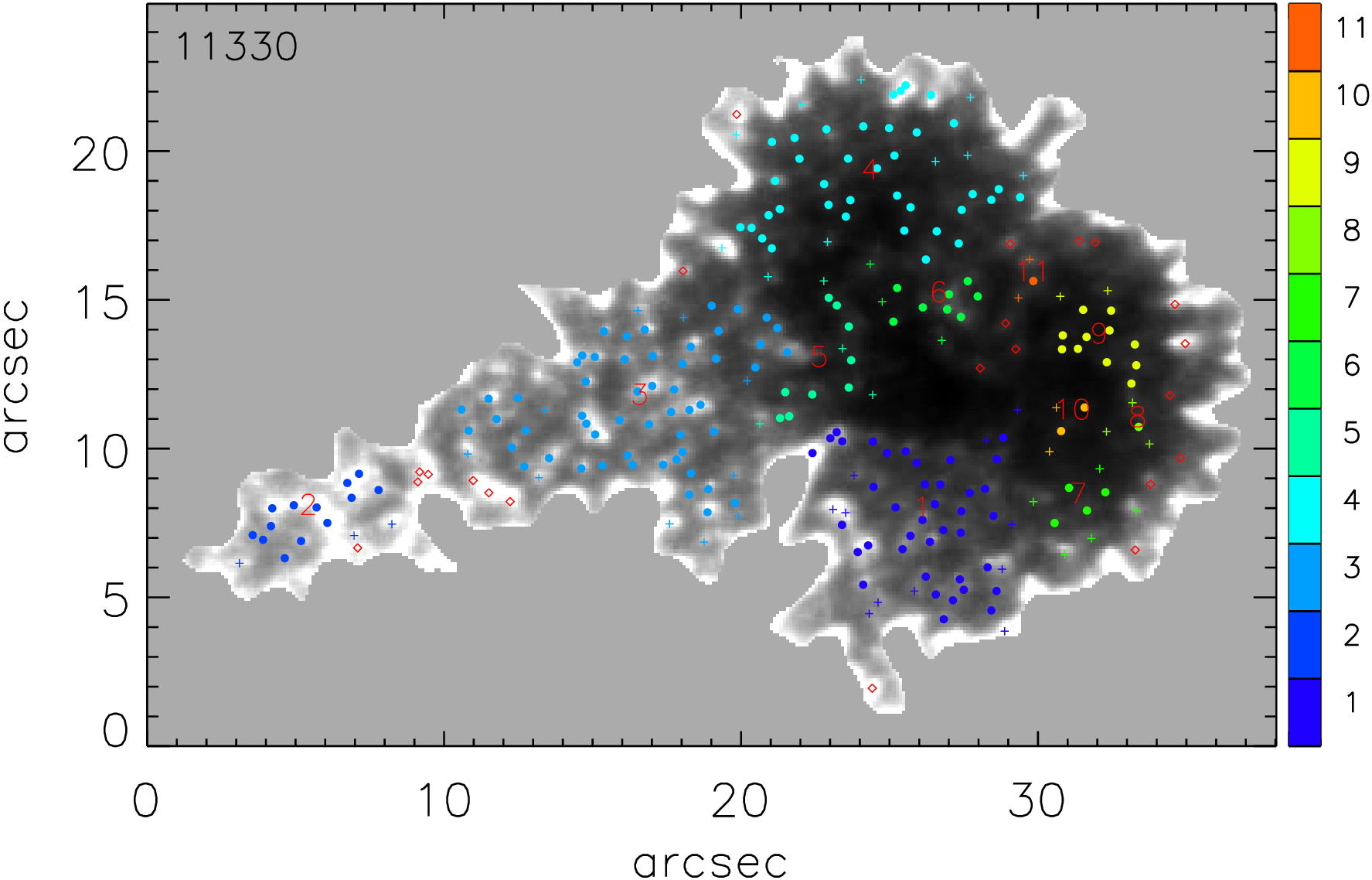}
\hspace{-5pt}
\includegraphics[angle = 90,width=0.5\textwidth]{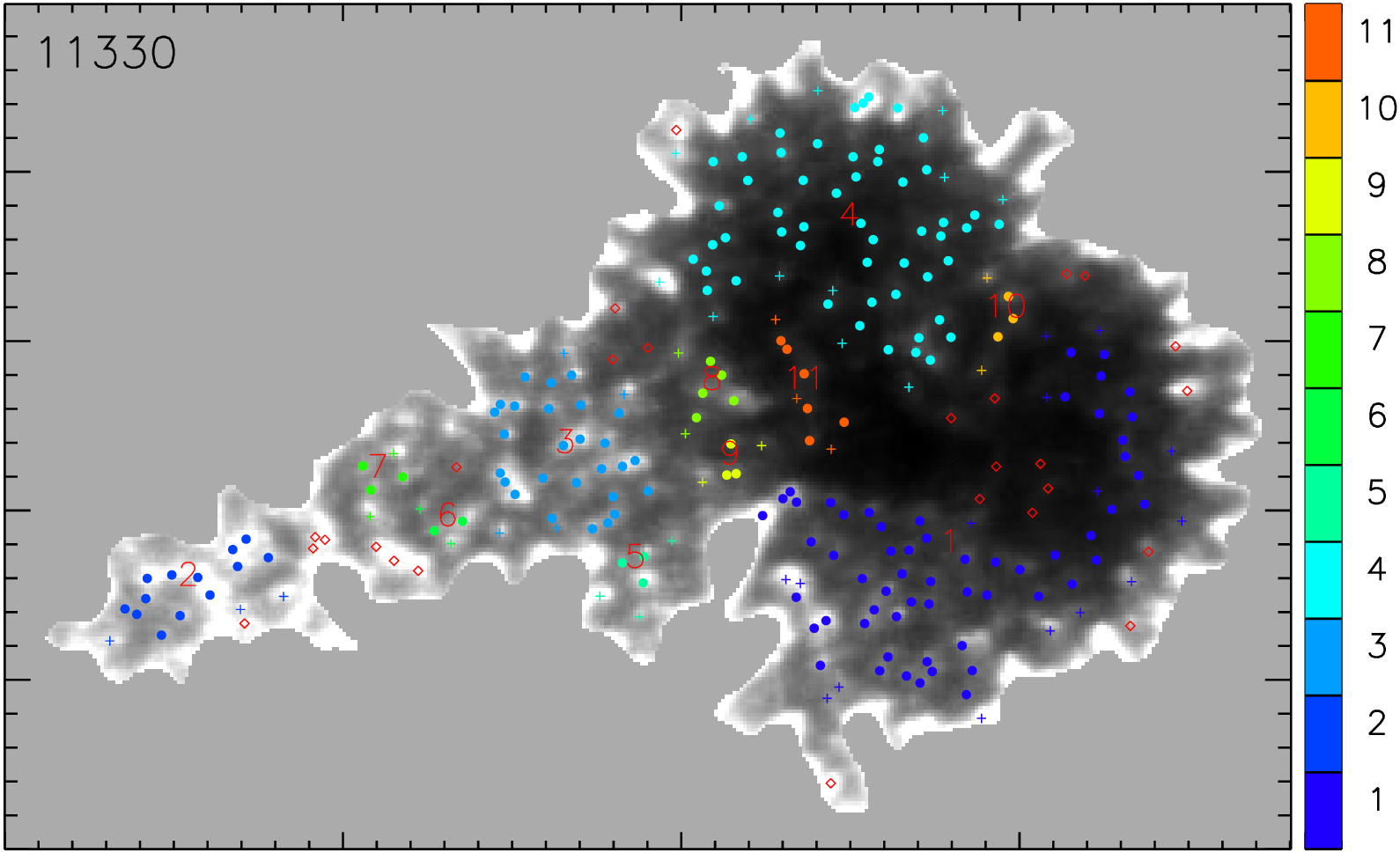}
}
\vspace{10pt}
\caption{Same as Fig.~\ref{fig09} but for AR 10953 (top) and AR 11330 (bottom).}
\label{fig10}
\end{figure*}

\subsection{Density-based Clustering of UDs}
\label{cluster}
In this section, we look at the differences in the clustering of UDs identified by the two MLT schemes. As discussed in 
Sect.~\ref{dbscan}, DBSCAN requires two parameters, namely, $\epsilon$ and $M$ to cluster the data into core, border, and noise points.
In order to choose the optimal combination of these two parameters, the routine was run by varying $\epsilon$ from 10 to 20\,pixels in steps 
of 0.1\,pixel and $M$ from 4 to 10\,points in steps of 1\,unit. The results obtained from this parameter space are shown in Figs.~\ref{fig07} 
and \ref{fig08}. It is seen that the number of clusters decreases monotonically with increasing $\epsilon$ (top panel) when the value of $M$ 
is between 4 and 6\,points. For higher $M$ values, the number of clusters increases and then decreases with increasing $\epsilon$ values, with 
the maximum number of clusters reducing with increasing $M$ values. The bottom panels of the figures show that the number of noise points 
monotonically decreases with $\epsilon$, while lower values of $M$ have a smaller number of noise points compared to higher $M$ values for 
a given $\epsilon$ value. The dotted line in the bottom panel indicates the 10\% level of all data points for the sunspot, which is used 
to ascertain the optimum value of $\epsilon$ and $M$ for which there are a sufficient number of clusters, which is also verified visually.
With this approach, we obtain a range of $\epsilon$ values between 10.5 and 11.5\,pixels 
and $M$ between 4 and 5\,points. The choice of $\epsilon$ and $M$ values produces nearly identical results in the number of clusters as well 
as noise points for the regular and modified MLT approach. This is illustrated in Figs.~\ref{fig09} and \ref{fig10} where the colors 
indicate the cluster number. The number of clusters obtained for NOAA AR 10923, 10933, 10953, and 11330 are 18(17), 11(12), 
10(11), and 11(11) for the regular(modified) MLT scheme, respectively. Similarly, the number of noise points for the four sunspots are 55(57), 
43(41), 19(26), and 22(26) for the regular(modified) MLT scheme, respectively.

Figure~\ref{fig09} shows that there is an inhomogeneous distribution of UDs with a preferential clustering in certain regions
or locations in the umbra versus others. For instance, in AR 10923, clusters 2 and 6 (top-left panel) and clusters 1 and 5 (top-right panel) are 
located on opposite sides of the umbra along the major axis of the spot. Clusters 2 (left) and 1 (right) also include a light bridge and other 
UDs in its vicinity. Clusters 1 (bottom left) and 2 (bottom right) in AR 10933 dominate the western side of the sunspot, while clusters 6 
(bottom left) and 7 (bottom right) are co-spatial with a light bridge. A similar trend is seen in AR 10953 (top panels of Fig.~\ref{fig10}) 
with clusters 1, 6, and 8 using both MLT methods, while cluster 2 is exclusively located in the smaller umbral core in the southeastern part 
of the spot. The preferential clustering of UDs in the narrower regions of the umbra is also seen in AR 11330; however, in the regular method 
(bottom-left panel), cluster 3 occupies nearly the entire eastern core with a smaller contribution from cluster 2, while in the modified method,
the same region comprises seven smaller clusters, namely, 2, 3, 5, 6, 7, 8, and 9. On the other hand, the clustering in the larger umbral core exhibits
a reversed trend where the regular method is associated with nine clusters (bottom-left panel) while the modified method shows only four clusters.
It is also seen that the fraction of core and border points with more than 10 UDs in a cluster accounts for more than 70\% of all UDs. The number of clusters
where there are more than 10 UDs in each cluster varies from four in AR 11330 to eight in AR 10923. The fraction of clusters with 10 UDs or more shows a
linear correlation coefficient of 0.49 and 0.89 with the sunspot area for the regular and modified MLT approach, respectively. The noise points (red diamonds)
are mostly distributed near the umbra-penumbra boundary.

\section{Discussion}
\label{discuss}
In this article, we describe two approaches to the MLT method, which is often utilized in the detection of sunspot UDs, and ascertain the 
resulting differences in the physical quantities arising from the methods. The two approaches differ in the contour level within a 
patch that encloses a UD -- in the first regular method, the level is set at the average of the maximum intensity and background intensity of 
the patch, i.e., the full-width-at-half-maximum, 
while in the second modified method, the level is set at the average of the maximum intensity and mean intensity of the patch. In the 
modified approach, there can be multiple contours that enclose pixels with a local intensity maxima, while in the regular approach, there is only 
one contour. While both approaches differ in the contour level and the number of contours within an MLT patch, we also restrict the points 
of local intensity maxima as those having a value greater than 30\% of the minimum umbral intensity and less than the mean intensity of all 
remaining pixels of local maxima. As a consequence of the above, the median effective diameter from the regular approach is about 70--90\,km greater 
than that from the modified approach, where the median values range from 320 to 410\,km with the regular method. The larger 
effective diameter retrieved by the regular approach also yields a fill fraction that is around 6\% higher on average for all sunspots, as compared 
to those retrieved from the modified approach. The differences in the effective diameter are, however, within the spatial sampling of the Hinode blue continuum
images and also consistent with earlier quoted values 
\citep{1997A&A...328..682S,2002A&A...388.1048T,2009A&A...504..575S,2009ApJ...694.1080S,2009ApJ...702.1048W,2012ApJ...752..109L,2012ApJ...757...49W}.
The values of the fill fraction from both methods are also in agreement with previously reported values
\citep{1997A&A...328..682S,2002A&A...388.1048T,2018ApJ...855....8Y}.
On the other hand, the histograms of the mean intensity are nearly identical for both approaches, with 
only AR 10933 showing a small difference in the amplitude. Similarly, the density-based, spatial clustering of UDs from both approaches yields very 
similar results in terms of the number of clusters and noise points. The $\epsilon$ and $M$ values used in the clustering routine for both methods 
are within two pixels. Thus, the inclusion of points of local maxima along with MLT does not produce significant differences statistically.

We finally comment on the spatial clustering of UDs in the context of MHD simulations of magneto-convection in sunspots. 
\cite{2006ApJ...641L..73S} showed that the $\tau=1$ surface is elevated by around 300\,km at the location of the convective plume, corresponding to the UD
where the upflows are the strongest and the magnetic field is significantly reduced in comparison to the surroundings. However, the contrast in the velocity 
and magnetic signatures of the UDs is significantly reduced with respect to the adjacent umbra as one goes higher to $\tau=0.1$ and $0.01$, where most 
spectral lines form. The typical vertical extension of UDs varies between 0.6 and 0.9\,Mm during their temporal evolution \citep{2006ApJ...641L..73S}. 
While UDs are shallow and mostly, sub-photospheric phenomena, their spatial distribution within the umbra depends on the field strength and in 
particular its vertical component, which dictates whether a fluid parcel will be convectively unstable  
\citep{1966MNRAS.133...85G}. This is particularly evident in numerical simulations of a pair of opposite-polarity
sunspots adjacent to each other with field strengths of 3 and 4\,kG \citep{2009Sci...325..171R,2011ApJ...729....5R}, where 
the inhomogeneous clustering of UDs in the sunspot with a lower field strength is consistent with our results shown in Figs.~\ref{fig09} and ~\ref{fig10}.
A point to be noted is that while radiative MHD simulations provide three-dimensional cubes of physical quantities such as the vector magnetic field,
velocity, temperature, pressure, and density, etc., they do not provide a mask for the locations of UDs, which would permit an absolute ground truth to compare with 
observations. In addition, the sunspot fine structure in the visible photospheric layers where spectral lines typically form is sensitive to the boundary conditions used
in the MHD framework that can produce significant variations in the spatial and photometric distribution of UDs, as described above. Thus, while  
comparisons between different tracking methods are possible, ascertaining a superior routine is possible only when the quantities are known a priori.
The preferential regions in the umbra, where UDs are populous, would suggest that the classical categorization of UDs as peripheral and central, based 
on their proximity to the umbra-penumbra boundary, requires a reconsideration. While we have presented only four sunspots in this article, 
we plan to extend our clustering analysis to a larger sample of spots at different evolutionary stages, in-situ conditions, morphologies, and magnetic 
complexities. This would permit a better understanding of the conditions that lead to a localization of convective intrusions within a sunspot umbra
that can be compared with numerical simulations.

\section{Conclusions}
\label{conclu}
The modification of the standard MLT routine, used in the detection of sunspot UDs, to include local intensity enhancements produces
a smaller effective diameter within the range of 70--90\,km and a reduced fill fraction of around 6\% in comparison to the regular scheme. However, these differences
are still within the range of values cited by earlier works. 
On the other hand, the distribution of the mean intensity of UDs from both methods is nearly identical.
A density-based spatial clustering of the UDs also shows similar results with both detection routines. In addition, there is a preferential clustering of UDs 
in the umbra, particularly in the vicinity of faint light bridges or where the umbral core is constricted. 
Further analysis is required to ascertain the dependency of the clustering of UDs with the in-situ and external environment of the sunspot 
and include the clustering to time-dependent quantities such as horizontal speeds and lifetimes, etc.
An absolute ground truth is essential to clarify which tracking routine is superior; however the unique intrinsic 
magnetic environment of an AR, renders such a distinction, non-trivial.

\begin{acknowledgments}
\textit{Hinode} is a Japanese mission developed and launched by
ISAS/JAXA, collaborating with NAOJ as a domestic partner,
NASA and STFC (UK) as international partners. Scientific
operation of the Hinode mission is conducted by the \textit{Hinode}
science team organized at ISAS/JAXA. This team mainly
consists of scientists from institutes in the partner countries.
Support for the post-launch operation is provided by JAXA and
NAOJ (Japan), STFC (UK), NASA, ESA, and NSC (Norway).
We thank the referee for providing insightful and useful comments.
\end{acknowledgments}

\begin{contribution}
R.E.L formulated the idea, compiled the MLT and DBSCAN routines in IDL, performed the analysis, and wrote the manuscript. A.C. selected the \textit{Hinode} 
observations and reduced the data. 
\end{contribution}


\bibliography{louis_ref}{}

@ARTICLE{1993PatRe..26..375B,
       author = {{Barrodale}, I. and {Skea}, D. and {Berkley}, M. and {Kuwahara}, R. and {Poeckert}, R.},
        title = "{Warping digital images using thin plate splines}",
      journal = {Pattern Recognition},
         year = 1993,
        month = jan,
       volume = {26},
       number = {2},
        pages = {375-376},
          doi = {10.1016/0031-3203(93)90045-X},
       adsurl = {https://ui.adsabs.harvard.edu/abs/1993PatRe..26..375B}
}

@ARTICLE{1968SoPh....4..303B,
       author = {{Beckers}, J.~M. and {Schr{\"o}ter}, E.~H.},
        title = "{The Intensity, Velocity and Magnetic Structure of a Sunspot Region. II: Some Properties of Umbral Dots}",
      journal = {\solphys},
         year = 1968,
        month = jul,
       volume = {4},
       number = {3},
        pages = {303-314},
          doi = {10.1007/BF00149561},
       adsurl = {https://ui.adsabs.harvard.edu/abs/1968SoPh....4..303B}
}

@ARTICLE{2010A&A...510A..12B,
       author = {{Bharti}, L. and {Beeck}, B. and {Sch{\"u}ssler}, M.},
        title = "{Properties of simulated sunspot umbral dots}",
      journal = {\aap},
         year = 2010,
        month = jan,
       volume = {510},
          eid = {A12},
        pages = {A12},
          doi = {10.1051/0004-6361/200913328},
archivePrefix = {arXiv},
       eprint = {0911.5068},
 primaryClass = {astro-ph.SR},
       adsurl = {https://ui.adsabs.harvard.edu/abs/2010A&A...510A..12B}
}

@ARTICLE{1941VAG....76..194B,
       author = {{Biermann}, L.},
        title = "{Der gegenw{\"a}rtige Stand der Theorie konvektiver Sonnenmodelle}",
      journal = {Vierteljahresschrift der Astronomischen Gesellschaft},
         year = 1941,
        month = jan,
       volume = {76},
        pages = {194-200},
       adsurl = {https://ui.adsabs.harvard.edu/abs/1941VAG....76..194B}
}

@ARTICLE{2001SoPh..201...13B,
       author = {{Bovelet}, B. and {Wiehr}, E.},
        title = "{A New Algorithm for Pattern Recognition and its Application to Granulation and Limb Faculae}",
      journal = {\solphys},
         year = 2001,
        month = jun,
       volume = {201},
       number = {1},
        pages = {13-26},
          doi = {10.1023/A:1010344827952},
       adsurl = {https://ui.adsabs.harvard.edu/abs/2001SoPh..201...13B}
}

@ARTICLE{1963MNRAS.126..431C,
       author = {{Chitre}, S.~M.},
        title = "{The structure of sunspots}",
      journal = {\mnras},
         year = 1963,
        month = jan,
       volume = {126},
        pages = {431},
          doi = {10.1093/mnras/126.5.431},
       adsurl = {https://ui.adsabs.harvard.edu/abs/1963MNRAS.126..431C}
}

@ARTICLE{1986ApJ...302..809C,
       author = {{Choudhuri}, A.~R.},
        title = "{The Dynamics of Magnetically Trapped Fluids. I. Implications for Umbral Dots and Penumbral Grains}",
      journal = {\apj},
         year = 1986,
        month = mar,
       volume = {302},
        pages = {809},
          doi = {10.1086/164042},
       adsurl = {https://ui.adsabs.harvard.edu/abs/1986ApJ...302..809C}
}

@ARTICLE{1964ApJ...139...45D,
       author = {{Danielson}, Robert E.},
        title = "{The Structure of Sunspot Umbras. I. Observations.}",
      journal = {\apj},
         year = 1964,
        month = jan,
       volume = {139},
        pages = {45},
          doi = {10.1086/147738},
       adsurl = {https://ui.adsabs.harvard.edu/abs/1964ApJ...139...45D}
}

@ARTICLE{1965ApJ...141..548D,
       author = {{Deinzer}, Willi},
        title = "{On the Magneto-Hydrostatic Theory of Sunspots.}",
      journal = {\apj},
         year = 1965,
        month = feb,
       volume = {141},
        pages = {548},
          doi = {10.1086/148144},
       adsurl = {https://ui.adsabs.harvard.edu/abs/1965ApJ...141..548D}
}

@ARTICLE{1966MNRAS.133...85G,
       author = {{Gough}, D.~O. and {Tayler}, R.~J.},
        title = "{The influence of a magnetic field on Schwarzschild's criterion for convective instability in an ideally conducting fluid}",
      journal = {\mnras},
         year = 1966,
        month = jan,
       volume = {133},
        pages = {85},
          doi = {10.1093/mnras/133.1.85},
       adsurl = {https://ui.adsabs.harvard.edu/abs/1966MNRAS.133...85G}
}

@ARTICLE{1986A&A...156..347G,
       author = {{Grossmann-Doerth}, U. and {Schmidt}, W. and {Schroeter}, E.~H.},
        title = "{Size and temperature of umbral dots}",
      journal = {\aap},
         year = 1986,
        month = feb,
       volume = {156},
        pages = {347},
       adsurl = {https://ui.adsabs.harvard.edu/abs/1986A&A...156..347G}
}

@ARTICLE{2008SoPh..250...17H,
       author = {{Hamedivafa}, H.},
        title = "{Application of an Improved Method of Image Segmentation and Some Considerations on Identification and Tracking of Umbral Dots}",
      journal = {\solphys},
         year = 2008,
        month = jul,
       volume = {250},
       number = {1},
        pages = {17-29},
          doi = {10.1007/s11207-008-9201-0},
       adsurl = {https://ui.adsabs.harvard.edu/abs/2008SoPh..250...17H}
}

@ARTICLE{2007SoPh..243....3K,
       author = {{Kosugi}, T. and {Matsuzaki}, K. and {Sakao}, T. and {Shimizu}, T. and {Sone}, Y. and {Tachikawa}, S. and {Hashimoto}, T. and {Minesugi}, K. and {Ohnishi}, A. and {Yamada}, T. and {Tsuneta}, S. and {Hara}, H. and {Ichimoto}, K. and {Suematsu}, Y. and {Shimojo}, M. and {Watanabe}, T. and {Shimada}, S. and {Davis}, J.~M. and {Hill}, L.~D. and {Owens}, J.~K. and {Title}, A.~M. and {Culhane}, J.~L. and {Harra}, L.~K. and {Doschek}, G.~A. and {Golub}, L.},
        title = "{The Hinode (Solar-B) Mission: An Overview}",
      journal = {\solphys},
         year = 2007,
        month = jun,
       volume = {243},
       number = {1},
        pages = {3-17},
          doi = {10.1007/s11207-007-9014-6},
       adsurl = {https://ui.adsabs.harvard.edu/abs/2007SoPh..243....3K}
}

@ARTICLE{1979A&A....79..128L,
       author = {{Loughhead}, R.~E. and {Bray}, R.~J. and {Tappere}, E.~J.},
        title = "{Improved observations of sunspot umbral dots.}",
      journal = {\aap},
         year = 1979,
        month = oct,
       volume = {79},
       number = {1-2},
        pages = {128-131},
       adsurl = {https://ui.adsabs.harvard.edu/abs/1979A&A....79..128L}
}

@ARTICLE{2012ApJ...752..109L,
       author = {{Louis}, Rohan E. and {Mathew}, Shibu K. and {Bellot Rubio}, Luis R. and {Ichimoto}, Kiyoshi and {Ravindra}, B. and {Raja Bayanna}, A.},
        title = "{Properties of Umbral Dots from Stray Light Corrected Hinode Filtergrams}",
      journal = {\apj},
         year = 2012,
        month = jun,
       volume = {752},
       number = {2},
          eid = {109},
        pages = {109},
          doi = {10.1088/0004-637X/752/2/109},
archivePrefix = {arXiv},
       eprint = {1204.4088},
 primaryClass = {astro-ph.SR},
       adsurl = {https://ui.adsabs.harvard.edu/abs/2012ApJ...752..109L}
}

@ARTICLE{2024AdSpR..73.3256L,
       author = {{Louis}, Rohan Eugene and {Mathew}, Shibu K. and {Raja Bayanna}, A.},
        title = "{Classification of circular polarization Stokes profiles in a sunspot using k-means clustering}",
      journal = {Advances in Space Research},
         year = 2024,
        month = mar,
       volume = {73},
       number = {6},
        pages = {3256-3269},
          doi = {10.1016/j.asr.2023.12.046},
archivePrefix = {arXiv},
       eprint = {2401.03908},
 primaryClass = {astro-ph.SR},
       adsurl = {https://ui.adsabs.harvard.edu/abs/2024AdSpR..73.3256L}
}

@ARTICLE{1769Meister,
       author = {{Meister}, A.~L.~F.},
        title = "{Generalia de genesi figurarum planarum et inde pendentibus earum affectionibus}",
      journal = {Nov. Com. G\"ott.},
         year = 1769,
       volume = {1},
        pages = {144},
          doi = {},
archivePrefix = {},
       eprint = {},
 primaryClass = {},
       adsurl = {}
}

@ARTICLE{2010ApJ...713.1282O,
       author = {{Ortiz}, A. and {Bellot Rubio}, L.~R. and {Rouppe van der Voort}, L.},
        title = "{Downflows in Sunspot Umbral Dots}",
      journal = {\apj},
         year = 2010,
        month = apr,
       volume = {713},
       number = {2},
        pages = {1282-1291},
          doi = {10.1088/0004-637X/713/2/1282},
archivePrefix = {arXiv},
       eprint = {1003.1897},
 primaryClass = {astro-ph.SR},
       adsurl = {https://ui.adsabs.harvard.edu/abs/2010ApJ...713.1282O}
}

@ARTICLE{1979ApJ...234..333P,
       author = {{Parker}, E.~N.},
        title = "{Sunspots and the physics of magnetic flux tubes. IX. Umbral dots and longitudinal overstability.}",
      journal = {\apj},
         year = 1979,
        month = nov,
       volume = {234},
        pages = {333-347},
          doi = {10.1086/157501},
       adsurl = {https://ui.adsabs.harvard.edu/abs/1979ApJ...234..333P}
}

@ARTICLE{2011ApJ...729....5R,
       author = {{Rempel}, M.},
        title = "{Penumbral Fine Structure and Driving Mechanisms of Large-scale Flows in Simulated Sunspots}",
      journal = {\apj},
         year = 2011,
        month = mar,
       volume = {729},
       number = {1},
          eid = {5},
        pages = {5},
          doi = {10.1088/0004-637X/729/1/5},
archivePrefix = {arXiv},
       eprint = {1101.2200},
 primaryClass = {astro-ph.SR},
       adsurl = {https://ui.adsabs.harvard.edu/abs/2011ApJ...729....5R}
}

@ARTICLE{2009Sci...325..171R,
       author = {{Rempel}, M. and {Sch{\"u}ssler}, M. and {Cameron}, R.~H. and {Kn{\"o}lker}, M.},
        title = "{Penumbral Structure and Outflows in Simulated Sunspots}",
      journal = {Science},
         year = 2009,
        month = jul,
       volume = {325},
       number = {5937},
        pages = {171},
          doi = {10.1126/science.1173798},
archivePrefix = {arXiv},
       eprint = {0907.2259},
 primaryClass = {astro-ph.SR},
       adsurl = {https://ui.adsabs.harvard.edu/abs/2009Sci...325..171R}
}

@ARTICLE{2008A&A...492..233R,
       author = {{Riethm{\"u}ller}, T.~L. and {Solanki}, S.~K. and {Zakharov}, V. and {Gandorfer}, A.},
        title = "{Brightness, distribution, and evolution of sunspot umbral dots}",
      journal = {\aap},
         year = 2008,
        month = dec,
       volume = {492},
       number = {1},
        pages = {233-243},
          doi = {10.1051/0004-6361:200810701},
archivePrefix = {arXiv},
       eprint = {0812.0477},
 primaryClass = {astro-ph},
       adsurl = {https://ui.adsabs.harvard.edu/abs/2008A&A...492..233R}
}

@ARTICLE{2008ApJ...672..684R,
       author = {{Rimmele}, T.},
        title = "{On the Relation between Umbral Dots, Dark-cored Filaments, and Light Bridges}",
      journal = {\apj},
         year = 2008,
        month = jan,
       volume = {672},
       number = {1},
        pages = {684-695},
          doi = {10.1086/523702},
       adsurl = {https://ui.adsabs.harvard.edu/abs/2008ApJ...672..684R}
}

@ARTICLE{2019ApJ...875L..18S,
       author = {{Sainz Dalda}, Alberto and {de la Cruz Rodr{\'\i}guez}, Jaime and {De Pontieu}, Bart and {Go{\v{s}}i{\'c}}, Milan},
        title = "{Recovering Thermodynamics from Spectral Profiles observed by IRIS: A Machine and Deep Learning Approach}",
      journal = {\apjl},
         year = 2019,
        month = apr,
       volume = {875},
       number = {2},
          eid = {L18},
        pages = {L18},
          doi = {10.3847/2041-8213/ab15d9},
archivePrefix = {arXiv},
       eprint = {1904.08390},
 primaryClass = {astro-ph.SR},
       adsurl = {https://ui.adsabs.harvard.edu/abs/2019ApJ...875L..18S}
}

@ARTICLE{2006ApJ...641L..73S,
       author = {{Sch{\"u}ssler}, M. and {V{\"o}gler}, A.},
        title = "{Magnetoconvection in a Sunspot Umbra}",
      journal = {\apjl},
         year = 2006,
        month = apr,
       volume = {641},
       number = {1},
        pages = {L73-L76},
          doi = {10.1086/503772},
archivePrefix = {arXiv},
       eprint = {astro-ph/0603078},
 primaryClass = {astro-ph},
       adsurl = {https://ui.adsabs.harvard.edu/abs/2006ApJ...641L..73S}
}

@ARTICLE{1997A&A...328..689S,
       author = {{Sobotka}, Michal and {Brandt}, Peter N. and {Simon}, George W.},
        title = "{Fine structure in sunspots. II. Intensity variations and proper motions of umbral dots}",
      journal = {\aap},
         year = 1997,
        month = dec,
       volume = {328},
        pages = {689-694},
       adsurl = {https://ui.adsabs.harvard.edu/abs/1997A&A...328..689S}
}

@ARTICLE{1997A&A...328..682S,
       author = {{Sobotka}, Michal and {Brandt}, Peter N. and {Simon}, George W.},
        title = "{Fine structure in sunspots. I. Sizes and lifetimes of umbral dots}",
      journal = {\aap},
         year = 1997,
        month = dec,
       volume = {328},
        pages = {682-688},
       adsurl = {https://ui.adsabs.harvard.edu/abs/1997A&A...328..682S}
}

@ARTICLE{2009A&A...504..575S,
       author = {{Sobotka}, M. and {Puschmann}, K.~G.},
        title = "{Morphology and evolution of umbral dots and their substructures}",
      journal = {\aap},
         year = 2009,
        month = sep,
       volume = {504},
       number = {2},
        pages = {575-581},
          doi = {10.1051/0004-6361/200912365},
archivePrefix = {arXiv},
       eprint = {0907.4236},
 primaryClass = {astro-ph.SR},
       adsurl = {https://ui.adsabs.harvard.edu/abs/2009A&A...504..575S}
}

@ARTICLE{2009ApJ...694.1080S,
       author = {{Sobotka}, M. and {Jur{\v{c}}{\'a}k}, J.},
        title = "{Evolution of Physical Characteristics of Umbral Dots and Penumbral Grains}",
      journal = {\apj},
         year = 2009,
        month = apr,
       volume = {694},
       number = {2},
        pages = {1080-1084},
          doi = {10.1088/0004-637X/694/2/1080},
       adsurl = {https://ui.adsabs.harvard.edu/abs/2009ApJ...694.1080S}
}

@ARTICLE{2002A&A...388.1048T,
       author = {{Tritschler}, A. and {Schmidt}, W.},
        title = "{Sunspot photometry with phase diversity. II. Fine-structure characteristics}",
      journal = {\aap},
         year = 2002,
        month = jun,
       volume = {388},
        pages = {1048-1061},
          doi = {10.1051/0004-6361:20020542},
       adsurl = {https://ui.adsabs.harvard.edu/abs/2002A&A...388.1048T}
}

@ARTICLE{2008SoPh..249..167T,
       author = {{Tsuneta}, S. and {Ichimoto}, K. and {Katsukawa}, Y. and {Nagata}, S. and {Otsubo}, M. and {Shimizu}, T. and {Suematsu}, Y. and {Nakagiri}, M. and {Noguchi}, M. and {Tarbell}, T. and {Title}, A. and {Shine}, R. and {Rosenberg}, W. and {Hoffmann}, C. and {Jurcevich}, B. and {Kushner}, G. and {Levay}, M. and {Lites}, B. and {Elmore}, D. and {Matsushita}, T. and {Kawaguchi}, N. and {Saito}, H. and {Mikami}, I. and {Hill}, L.~D. and {Owens}, J.~K.},
        title = "{The Solar Optical Telescope for the Hinode Mission: An Overview}",
      journal = {\solphys},
         year = 2008,
        month = jun,
       volume = {249},
       number = {2},
        pages = {167-196},
          doi = {10.1007/s11207-008-9174-z},
archivePrefix = {arXiv},
       eprint = {0711.1715},
 primaryClass = {astro-ph},
       adsurl = {https://ui.adsabs.harvard.edu/abs/2008SoPh..249..167T}
}

@ARTICLE{2011A&A...530A..14V,
       author = {{Viticchi{\'e}}, B. and {S{\'a}nchez Almeida}, J.},
        title = "{Asymmetries of the Stokes V profiles observed by HINODE SOT/SP in the quiet Sun}",
      journal = {\aap},
         year = 2011,
        month = jun,
       volume = {530},
          eid = {A14},
        pages = {A14},
          doi = {10.1051/0004-6361/201016096},
archivePrefix = {arXiv},
       eprint = {1103.1987},
 primaryClass = {astro-ph.SR},
       adsurl = {https://ui.adsabs.harvard.edu/abs/2011A&A...530A..14V}
}

@ARTICLE{2018ApJ...855....8Y,
       author = {{Yadav}, Rahul and {Louis}, Rohan E. and {Mathew}, Shibu K.},
        title = "{Investigating the Relation between Sunspots and Umbral Dots}",
      journal = {\apj},
         year = 2018,
        month = mar,
       volume = {855},
       number = {1},
          eid = {8},
        pages = {8},
          doi = {10.3847/1538-4357/aaaeba},
archivePrefix = {arXiv},
       eprint = {1802.05088},
 primaryClass = {astro-ph.SR},
       adsurl = {https://ui.adsabs.harvard.edu/abs/2018ApJ...855....8Y}
}

@ARTICLE{2009ApJ...702.1048W,
       author = {{Watanabe}, H. and {Kitai}, R. and {Ichimoto}, K.},
        title = "{Characteristic Dependence of Umbral Dots on Their Magnetic Structure}",
      journal = {\apj},
         year = 2009,
        month = sep,
       volume = {702},
       number = {2},
        pages = {1048-1057},
          doi = {10.1088/0004-637X/702/2/1048},
archivePrefix = {arXiv},
       eprint = {0907.2750},
 primaryClass = {astro-ph.SR},
       adsurl = {https://ui.adsabs.harvard.edu/abs/2009ApJ...702.1048W}
}

@ARTICLE{2012ApJ...757...49W,
       author = {{Watanabe}, Hiroko and {Bellot Rubio}, Luis R. and {de la Cruz Rodr{\'\i}guez}, Jaime and {Rouppe van der Voort}, Luc},
        title = "{Temporal Evolution of Velocity and Magnetic Field in and around Umbral Dots}",
      journal = {\apj},
         year = 2012,
        month = sep,
       volume = {757},
       number = {1},
          eid = {49},
        pages = {49},
          doi = {10.1088/0004-637X/757/1/49},
archivePrefix = {arXiv},
       eprint = {1207.6006},
 primaryClass = {astro-ph.SR},
       adsurl = {https://ui.adsabs.harvard.edu/abs/2012ApJ...757...49W}
}

@ARTICLE{1990MNRAS.245..434W,
       author = {{Weiss}, N.~O. and {Brownjohn}, D.~P. and {Hurlburt}, N.~E. and {Proctor}, M.~R.~E.},
        title = "{Oscillatory convection in sunspot umbrae}",
      journal = {\mnras},
         year = 1990,
        month = aug,
       volume = {245},
       number = {3},
        pages = {434-434},
          doi = {10.1093/mnras/245.3.434},
       adsurl = {https://ui.adsabs.harvard.edu/abs/1990MNRAS.245..434W}
}

@ARTICLE{otsu_imgseg,
  author={Otsu, Nobuyuki},
  journal={IEEE Transactions on Systems, Man, and Cybernetics}, 
  title={A Threshold Selection Method from Gray-Level Histograms}, 
  year={1979},
  volume={9},
  number={1},
  pages={62-66},
  doi={10.1109/TSMC.1979.4310076}
}

@article{liao2001fast,
  title={A fast algorithm for multilevel thresholding},
  author={Liao, Ping-Sung and Chen, Tse-Sheng and Chung, Pau-Choo and others},
  journal={J. Inf. Sci. Eng.},
  volume={17},
  number={5},
  pages={713--727},
  year={2001}
}
\bibliographystyle{aasjournalv7}

\end{document}